\documentstyle[preprint,aps]{revtex}

\def\coeff#1#2{{\textstyle{#1\over #2}}}
\def\half{\coeff12 }

\def\beq{\begin{equation}}
\def\eeq{\end{equation}}
\def\bra#1{\langle #1 |}
\def\ket#1{| #1 \rangle}
\def\ov{\overline}

\def\frak{\bf}

\makeatletter
\def\theequation{\arabic{section}.\theequation@prefix\arabic{equation}}%
\def\mathletters{%
\inc@eqnnum  \setcounter{eqletter}{0}%
\edef\@currentlabel{\theequation}%
\def\theequation{\arabic{section}.\theequation@prefix\arabic{equation}%
           \alph{eqletter}}%
\def\inc@eqnnum{\addtocounter{eqletter}{1}}%
\def\dec@eqnnum{\addtocounter{eqletter}{-1}}%
}
\def\section{{\setcounter{equation}{0}}
\@mainheadtrue
\@startsection {section}{1}{\z@}{0.8cm plus1ex minus
 .2ex}{0.5cm plus1ex minus.2ex}{\reset@font\small\bf\centering}
           }
\makeatother

\begin{document}

\def\footnoterule{\hrule width \hsize}
\skip\footins = 12pt
\footskip     = 18pt
\footnotesep  = 10pt


\textwidth=6.5in
\hsize=6.5in
\oddsidemargin=0in
\evensidemargin=0in
\hoffset=0in

\textheight=9.5in
\vsize=9.5in
\topmargin=-.5in
\voffset=-.3in

\title{Physical States in Matter-Coupled Dilaton Gravity\footnotemark[1]}

\def\footstrut{\baselineskip 16pt}

\footnotetext[1]
{\baselineskip=16pt
This work is supported in part
by funds provided by
the U.S.\ National Science Foundation (N.S.F.) under contract
\#PHY-92-18990 (DC)
and in part by
the U.S.~Department of Energy (D.O.E.)
under contract \#DE-FC02-94ER40818 (RJ and BZ). \hfil\break
UCLA/95/TEP/16, MIT-CTP-2436 \hfil
Submitted to: {\em Annals Of Physics\/} \hfil
May 1995\break}

\author{D. Cangemi}
\medskip
\address{Department of Physics,
University of California, Los Angeles\\[-1ex]
405 Hilgard Ave., Los Angeles, CA ~90095--1547}

\bigskip

\author{R. Jackiw \\ and \\ B. Zwiebach}

\address{Center for Theoretical Physics,
Laboratory for Nuclear Science
and Department of Physics \\[-1ex]
Massachusetts Institute of Technology, Cambridge, MA ~02139--4307}

\maketitle

\setcounter{page}{0}
\thispagestyle{empty}

\begin{abstract}%
\noindent%
\baselineskip=16pt%
We revisit the quantization of matter-coupled, two-dimensional dilaton
gravity.  At the classical level and with a cosmological term, a series
of field transformations leads to a set of free fields of indefinite
signature. Without matter the system is represented by two scalar fields
of opposite signature.  With a particular quantization for the scalar
with negative kinetic energy, the system has zero central charge and we
find some physical states satisfying {\it all} the Virasoro conditions.
With matter, the constraints cannot be solved because of the Virasoro
anomaly. We discuss two avenues for consistent quantization:
modification of the constraints, and BRST quantization. The first avenue
appears to lead to very few physical states. The second, which roughly
corresponds to satisfying half of the Virasoro conditions, results in a
rich spectrum of physical states.  This spectrum, however, differs
significantly
from
that of free matter fields propagating on flat two-dimensional
space-time.
\end{abstract}

\newpage

\widetext

\section{Introduction and Summary}
\label{sec:1}

In the last few years much research has been carried out on models of
gravity in two space-time dimensions --- lineal gravity --- with the
hope of illuminating some puzzles of physical quantum gravity in four
dimensions: the problem of time, the problem of quantizing a
diffeomorphism invariant theory, the paradoxes associated with the
occurrence of black holes in the classical theory and Hawking
radiation in the semi-classical analysis, {\it etc\/}.

When modeling lineal gravity, two-dimensional analogs of Einstein's
equation and of the Hilbert-Einstein action cannot be used, because
they are vacuous in that dimensionality; therefore gravitational
dynamics has to be invented afresh.  The models that have been studied
recently posit local dynamics for the ``gravity'' sector, which
involves metric variables (metric tensor $g_{\mu \nu}\/$,
{\it Zweibein\/} $e^a_\mu\/$, spin-connection $\omega_\mu\/$) and
an additional world scalar (``dilaton'' or Lagrange multiplier field).
Such ``scalar-tensor'' theories, introduced a decade ago \cite{ref:1},
are obtained by dimensional reduction from higher-dimensional
Einstein theory \cite{ref:1,ref:1.5}.
They should be contrasted with models where quantum
fluctuations of matter variables induce gravitational
dynamics~\cite{ref:2}, which therefore are nonlocal and do not
appear to offer any insights into the questions posed by the physical,
four-dimensional theory.

Another lineal gravity theory is provided by two-dimensional string
theory, which is constructed in a rather indirect fashion as a field
theory on a background two-dimensional space-time.  A manifestly
background-independent formulation is not yet known, in contrast to the
case for the simpler field theories to be discussed here.  Therefore the
issue of summing over space-time metrics is unclear.  One would expect
the semi-classical limit of two-dimensional string theory to describe,
roughly speaking, nontrivial dynamics for a massless scalar field on a
two-dimensional background space-time, lacking translational invariance
\cite{ref:ginspargmoore}.

The model we study is the so called ``string-inspired dilaton
gravity'' -- CGHS theory ~\cite{ref:3}.  Initially
presented in variables that arise
naturally from string theory, the dilaton gravity action reads
\begin{equation}
\overline{I}_{\rm gravity} = \int d^2 x  \sqrt{- \overline{g}} \, e^{- 2 \phi}
           \left( \overline{R} + 4 \overline{g}^{\mu \nu}
           \partial_\mu \phi \partial_\nu \phi - \lambda \right)\,\,,
\label{eq:1:2}
\end{equation}
where $\phi\/$ is the dilaton field, $\overline{R}\/$ is the scalar
curvature constructed from the metric tensor $\overline{g}_{\mu \nu}
\, (\det \{ \overline{g}_{\mu \nu} \} \equiv \overline{g} )\/$ and
$\lambda\/$ is a cosmological constant.  The over-bar distinguishes
metric variables in the initial formulation from rescaled ones,
which we use henceforth,
that are defined by
\begin{mathletters}%
\label{eq:1:3all}%
\begin{eqnarray}
\eta  &=&  e^{-2 \phi} \,\,, \\
\label{eq:1:3a}
g_{\mu \nu}  &=&  e^{-2 \phi} \overline{g}_{\mu \nu}\,\,,
\label{eq:1:3b}
\end{eqnarray}%
\end{mathletters}%
in terms of which the action (\ref{eq:1:2}) becomes
\begin{equation}
I_{\rm gravity} = \int d^2 x \sqrt {-g} \,  ( \eta R - \lambda )\,\,.
\label{eq:1:4}
\end{equation}
We minimally couple a matter field $\varphi$,
\begin{equation}
I_{\rm matter} = \case{1}{2} \int d^2 x \sqrt{-g} \, g^{\mu \nu}
           \partial_\mu \varphi \partial_\nu \varphi\, ,
\label{eq:1:9}
\end{equation}
where, for simplicity we consider a single massless scalar field.
The coupling is conformally invariant and
consequently insensitive to the field redefinition~(\ref{eq:1:3b}).
Nevertheless, as is well known, quantum mechanically there is an anomaly.
(One may increase the number of matter fields $\varphi \to
\varphi^i\/$, $i = 1, \ldots, N\/$, but this is of no significance to
us; we do not consider the large-$N$ limit.)

The total action for the matter-coupled dilaton gravity
is the sum of (\ref{eq:1:4}) and (\ref{eq:1:9}),
weighted by the gravitational coupling constant~$G\/$,
\begin{equation}
I = \frac{1}{4 \pi G} I_{\rm gravity} + I_{\rm matter}\,\,.
\label{eq:1:10}
\end{equation}

We have previously analyzed this model within its equivalent,
Poincar\'{e} gauge invariant formulation~\cite{ref:9}.
Guided by its constraint structure, which in turn is a consequence of
gauge invariance, we solved some constraints and passed to new
variables by various redefinitions and canonical transformations,
arriving at variables that enter quadratically and decouple from each
other~\cite{ref:9}.  Our final Lagrange density reads
\begin{equation}
{\cal L}_{g + m}
           = \pi_a \dot{r}^a + \Pi \dot{\varphi} - u {\cal E} - v {\cal P}\,\,,
\label{eq:1:15}
\end{equation}
where $\{ \pi_a, r^a \}\/$ and $\{ u, v \}\/$ is what is left of the
``gravity'' variables and the remaining energy ${\cal E}\/$ and
momentum $\cal P\/$ constraints read
\begin{mathletters}%
\label{eq:1:16all}%
\begin{eqnarray}%
{\cal E}  &=&  -\case{1}{2}
           \left( \case{1}{\Lambda} \pi^a \pi_a
           + \Lambda {r^a}' r'_a \right)
           + \case{1}{2} \left( \Pi^2 + \varphi'^2 \right) \,\,,
           \label{eq:1:16a}  \\
{\cal P}  &=&  - \pi_a {r^a}' - \Pi \varphi'\,\,,
           \label{eq:1:16b}
\end{eqnarray}%
\end{mathletters}%
with $\Lambda \equiv \lambda / 8 \pi G\/$, and dot (dash) indicating
differentiation with respect to time $t$ (space $\sigma$).
Note that the dynamically active gravitational variables
$\left\{ \pi_a , r^a \right\}$ enter with an indefinite quadratic form,
regardless of sign $[\Lambda]$:
${1\over\Lambda} \pi^a \pi_a + \Lambda {r^a}' r'_a
= {1\over\Lambda} (\pi^0)^2
- {1\over\Lambda} (\pi^1)^2
+ \Lambda ({r^0}')^2
- \Lambda ({r^1}')^2$.
That there should be only two dynamical gravity variables
in this theory is consistent with the familiar observation that
two-dimensional dilaton
gravity involves the Liouville (Weyl factor) and dilaton degrees of freedom.

An important point should be made.
It is well-known \cite{ref:bilalcallan}
that a set of field redefinitions transforms the
dilaton gravity CGHS model without cosmological constant to
an indefinite sign quadratic form like in (\ref{eq:1:16a}).
The cosmological constant is then re-inserted as an approximate conformal
perturbation.
In our reduction,
involving field redefinitions that are local in time but not in space,
a quadratic indefinite form is the entire
expression, even in the presence of a cosmological constant.
As an
alternative to the previous gauge theoretic argument \cite{ref:9},
we shall present a derivation of
(\ref{eq:1:15}) and (\ref{eq:1:16all}) within the conventional
metric formalism \cite{ref:10}.

At this stage the development encounters a quantum obstruction: the
constraints that ${\cal E}\/$ and $\cal P\/$ should vanish appear
first class on the classical level --- they close under Poisson
bracketing --- but upon quantization they acquire a center and become
second class, owing to the well known triple-derivative Schwinger term
in the $[{\cal E, P}]\/$ commutator, which is the same as the Virasoro
anomaly.  The quantum theory
is recognized to be anomalous \cite{ref:9}. Much of our paper
deals with various issues concerning this
anomaly. Indeed, the central points in this paper concern
answers to the following questions:

\medskip
\noindent
(Q1)~ In pure dilaton gravity theory, {\it i.e.,} without matter fields,
the effective dynamics is that of two scalar fields of opposite signature.
In conformal field theory,
such a system is usually considered to have a non-vanishing central term
of $c=1+1=2$, with each scalar contributing unity, regardless of its
signature. If that is the case, how is it possible that there are solutions
to the constraints, as found in Refs.\cite{ref:9,ref:5} ?

\medskip
\noindent
(Q2)~ In string theory, one is not familiar with
states annihilated by {\it all} Virasoro operators. Typically, states
are annihilated by half of the Virasoro operators. Starting from
the wave functionals of Ref.\cite{ref:9}, one can construct the corresponding
states in the explicitly regulated language provided by an oscillator
expansion. Do these states satisfy {\it all} the Virasoro conditions?
How many such states are there?

\medskip
\noindent
(Q3)~ When matter is coupled, the constraints
have a center and cannot be solved. Is it possible to modify the
constraints and eliminate the center? What are then
the physical states? How does this compare with the BRST quantization
of the system?

\medskip
\noindent
(Q4)~ Is there a quantization scheme where the spectrum of physical
states in matter-coupled dilaton gravity is in rough agreement
with the spectrum of physical states
for free matter fields propagating in flat two-dimensional
space-time?

\medskip
\noindent
As a way to summarize our results, we now sketch the answers to
the above questions.

\medskip
\noindent
(A1)~ In the quantization scheme of Refs.\cite{ref:9,ref:5}
the  scalar with negative kinetic energy is treated in a  way different  from
what is common in conformal theory. When quantizing
such a scalar, one has the choice of either positive energy states
and negative norms, or negative energy states with
positive norms. For a scalar with positive kinetic term
one achieves both positive norms
and positive energies by identifying annihilation operators with positive
energy solutions and creation operators with negative energy solutions.
This choice is also the usual choice in conformal theory
for  scalars with negative kinetic energy; it
leads to positive energies, negative norms (thus the name ``negative
norm'' scalar) and a central charge of $(+1)$.
The opposite choice
of creation and annihilation operators leads to negative energies, positive
norms,  and a central charge of $(-1)$.
Therefore, there is a quantization scheme where
a scalar with positive kinetic energy and a scalar with negative
kinetic energy define a system with total
central charge zero.

\medskip
\noindent
(A2)~ We construct explicitly the Fock space representation of the two
states whose wave functionals are given in Refs.\cite{ref:9,ref:5}.
The two states
are exponentials of bilinears of creation operators acting
on the vacuum, differing by an overall sign in the exponential.
We have verified explicitly that they are annihilated by
all the Virasoro operators ${\cal L}_n$. Furthermore, we
find two more states that
are annihilated by all the ${\cal L}_n$'s, but curiously, they cannot be
represented by wave functionals.

\medskip
\noindent
(A3)~ We discuss two ways of modifying the constraints in order to obtain
a system with zero center and then we consider the alternative provided
by BRST quantization. In the first way to modify the constraints,
following some ideas of
K.~Kucha\v{r} \cite{ref:12}, we pass to yet further gravitational
variables and then find a simple addition that can be made to
${\cal E} \pm {\cal
P}\/$, which cancels the anomaly.  This addition is related to
structures that have appeared elsewhere in the
literature~\cite{ref:13}, and we explain this. Physical states
may exist but seem difficult to obtain. In the second way to modify
the constraints,
we follow standard conformal field theory lore and add a background
charge to the scalar with negative kinetic energy . This can be used to reduce
the central charge of this scalar to any desired value, and
in particular to the value that will give zero total central charge.
Despite having zero central charge, it seems unlikely that there are
many physical states. In the
second way, using the
BRST procedure, the ghost system is
coupled and it carries a center $(-26)$. Furthermore,
for the  positive-sign scalar
of the gravitational sector a background
charge is used to increase its center so that together with the
negative sign scalar [which here gives center $(+1)$] and matter, one
attains the center $(+26)$. Physical states are those states
that are annihilated by the BRST operator and cannot be written as the
BRST operator acting on another state. Roughly speaking, BRST quantization
is a consistent procedure to impose half of the constraints. The
spectrum of physical states for our present problem is rich and follows
from  well-known results \cite{ref:goldstone,ref:bilal}.

\medskip
\noindent
(A4)~ The procedure of modifying the constraints to get zero center, and
then imposing all of them seems unable to produce a large (or infinite)
number of physical states. In the BRST method the space-time is taken
to be a flat cylinder, and we
get a set of physical states that is not very different from that
one would expect in the zero gravity limit. The continuous parameters
that describe the states, however, do not work out. For a single
scalar field we should have one continuous parameter, the vacuum
``momentum'',
but here we obtain two continuous parameters. We feel that a large
mismatch between these sets of physical states is an indication that
the semi-classical approximation
to quantized gravity
may be problematic.

\medskip

\medskip
Let us describe briefly the organization of this paper.
In Section~\ref{sec:2} we review the gauge theoretical formulation
of the model.  In Section~\ref{sec:3} we explain how to
reduce the theory in terms of metric-based variables, to
indefinite sign quadratic form~\cite{ref:10}.
Section~\ref{sec:4} discusses
the modification of the constraints to achieve zero central term as in
Ref.\cite{ref:12}. In Section~\ref{sec:5}, we describe the alternative
quantizations of the  scalar field with negative kinetic energy.
We consider the
pure dilaton gravity theory
and give the oscillator description of the
physical states, showing that they satisfy all the Virasoro conditions.
In Section~\ref{sec:newsec} we show how a standard conformal improvement of the
negative norm scalar gives a constraint system with zero center.
In Section~\ref{sec:6} we describe
the BRST quantization of dilaton gravity with and without matter, and
compare the spectrum of physical states
to that of the flat-space limit. Finally, in Section~\ref{sec:8}
we offer some general comments, discuss open questions,
and speculate about the four-dimensional
theory, using the
insights drawn from the 2-dimensional toy model.

\section{Gauge Theoretical Starting Point}
\label{sec:2}

This Section has two parts. In the first part we
summarize the gauge theoretical formulation of the dilaton
gravity theory and  describe the constraint structure of the
theory. In the second part, for the benefit of the interested
reader, we give a self-contained derivation of most of the results
quoted in the first part.

\subsection{Summary of the Gauge Formulation}

The pure dilaton gravity theory given in (\ref{eq:1:4}), and
the matter-coupled dilaton gravity theory in (\ref{eq:1:9}),(\ref{eq:1:10})
can be given a gauge
theoretical formulation. The basic idea is to
work with Einstein-Cartan variables --- {\it Zweibein\/} and
spin-connection --- and to view them as gauge potentials for some
suitably chosen Lie group.  Specifically, one combines the
Einstein-Cartan variables into a Lie algebra valued gauge connection
$A_\mu\/$, builds the gauge curvature $F_{\mu \nu} = \partial_\mu
A_\nu - \partial_\nu A_\mu + [ A_\mu, A_\nu]\/$, introduces a
non-singular, bilinear invariant on the Lie algebra $\langle \; \mid
\; \rangle\/$, and forms the scalar-tensor action
\begin{equation}
I \propto \int \left< H \mid F_{\mu \nu} d x^\mu d x^\nu \right>\, ,
\label{eq:1:1}
\end{equation}
where $H\/$ comprises (a multiplet of) Lagrange multipliers.  (The
invariant bilinear $\langle \; \mid \; \rangle\/$ is constructed from
the Cartan-Killing metric when the group is semi-simple; otherwise
alternative expressions must be used.)  The dynamical equations
involve $H\/$ and $A_\mu\/$; when they are re-expressed in terms of
the dilaton field (which corresponds to one component of the $H\/$
multiplet) and in terms of the metric tensor (which is reconstructed
from the {\it Zweibein\/} and spin-connection that are collected in
$A_\mu\/$) one regains the metric version of the gravitational
equations.

The action (\ref{eq:1:1}) does not depend on a background metric, and
the hoped-for advantage in the gauge theoretical formulation of a
geometric gravity theory is that one can use quantization methods that
have been perfected during the decades spent studying gauge
theories, thereby circumventing some of the obstacles to quantizing a
gravity theory.  Moreover, enforcing gauge invariance can resolve
ambiguities and uncertainties in the quantization procedure.  Of
course, when the gauge theoretical approach succeeds, it should also
instruct, in retrospect, how the goal could have been reached if one
remained with the conventional formalism.

The theory given in (\ref{eq:1:4}) is formulated using the
centrally extended, 4-parameter Poincar\'{e} group~\cite{ref:4}.  In
the algebra of this group, there occur two translation generators $P_a$,
$(a = 0, 1)\/$ whose commutator contains the central element $I$,
\begin{mathletters}%
\label{eq:1:5all}%
\begin{equation}
[ P_a, P_b ] = \epsilon_{a b} I\,,
\label{eq:1:5a}
\end{equation}
while the commutator with the Lorentz rotation generator $J\/$ is unmodified.
\begin{equation}
[ P_a, J ] = \epsilon_{a}{}^{b} P_b\, .
\label{eq:1:5b}
\end{equation}%
\end{mathletters}%
[Notation: tangent space indices $(a, b, \ldots)\/$ are moved with
$h_{a b} \equiv {\rm diag} (1, -1)\/$; $\epsilon^{a b} = - \epsilon^{b
a}\/$, $\epsilon^{01} = 1\/$;
space-time points $x^\mu$ are $2$-vectors $(t,\sigma)$.]
The {\it Zweibein\/} $e^a_\mu\/$ is
taken to be the gauge potential for translations, the spin-connection
$\omega_\mu\/$ is the gauge potential for the Lorentz rotation, and a
further U(1) gauge potential $a_\mu\/$ has to be introduced for the
central element $I\/$.  Thus, the gauge connection is constructed as
\begin{mathletters}%
\label{eq:1:6all}%
\begin{equation}
A_\mu = e^a_\mu P_a + \omega_\mu J + a_\mu I\, ,
\label{eq:1:6a}
\end{equation}
and the curvature reads
\begin{eqnarray}
F  &=&  \case{1}{2} \epsilon^{\mu \nu} F_{\mu \nu}
           =  \epsilon^{\mu \nu}
           \left\{ \bigl( \partial_\mu e^a_\nu
                      + \epsilon^a{}_{b} \omega_\mu e^b_\nu \bigr)
           P_a + \partial_\mu \omega_\nu J
           + \bigl( \partial_\mu a_\nu
                      + \case{1}{2} e^a_\mu \epsilon_{a b} e^b_\nu  \bigr)
           I \right\} \, ,  \nonumber  \\
&\equiv& f^a P_a + f^2 J + f^3 I\,.
\label{eq:1:6b}
\end{eqnarray}%
\end{mathletters}%
Correspondingly, a quartet multiplet of Lagrange multiplier fields is
introduced
\begin{equation}
H = \eta_a h^{ab} P_b - \eta_3 J - \eta_2 I\,,
\label{eq:1:7}
\end{equation}
and the invariant, non-degenerate bilinear is given by
\begin{equation}
\langle H \mid F \rangle
           = \eta_a f^a + \eta_2 f^2 + \eta_3 f^3\,,
\label{eq:1:8}
\end{equation}
thus providing a Lagrange density $\cal L\/$ for the
action~(\ref{eq:1:1}).  It follows that the
dynamics implied by (\ref{eq:1:1})--(\ref{eq:1:8}) is the same as
that of (\ref{eq:1:4}) [or of (\ref{eq:1:2})]:
$\eta\/$ coincides with $ \frac{1}{2} \eta_2\/$,
$\sqrt{-g} \, R = 2 \epsilon^{\mu \nu} \partial_\mu \omega_\nu\/$
and the cosmological constant $\lambda\/$, absent from the gauge action,
emerges as the value in the solution for $\eta_3\/$, which
is also a group invariant~\cite{ref:4}.

The gauge theoretic formulation of the matter-coupled dilaton gravity theory
(\ref{eq:1:10}) requires further developments~\cite{ref:9}.  Gauge
invariance in the matter sector is achieved by a Higgs-like mechanism,
which introduces a new field $q^a\/$ --- the
{\it Poincar\'{e} coordinate\/} ---
and its canonical conjugate $p_a\/$~\cite{ref:8a,ref:8}.
The complete gauge invariant Lagrange density for the matter-gravity
system that we studied reads, in first order form~\cite{ref:9},
\begin{equation}
{\cal L}_{g + m} = \frac{1}{4 \pi G}
           \left( \eta_a \dot{e}^a_1 + \eta_2\dot{\omega}_1
           + \eta_3 \dot{a}_1  \right)
           + p_a \dot{q}^a + \Pi \dot{\varphi}
           + e^a_0 G_a + \omega_0 G_2 + a_0 G_3 - u {\cal E} - v {\cal P}\, ,
\label{eq:1:11}
\end{equation}
where,
\begin{mathletters}%
\label{eq:1:12all}%
\begin{eqnarray}
G_a  &\equiv& \frac{1}{4 \pi G} \left( \eta'_a
                      + \epsilon_a{}^{b} \eta_b \omega_1
           + \eta_3 \epsilon_{a b} e^b_1 \right)
           + \epsilon_a{}^{b} p_b\, ,
           \label{eq:1:12a}  \\
G_2  &\equiv& \frac{1}{4 \pi G} \left( \eta'_2
                      + \eta_a \epsilon^a{}_{b} e^b_1 \right)
           - q^a \epsilon_a{}^{b} p_b\, ,
           \label{eq:1:12b}  \\
G_3  &\equiv& \frac{1}{4 \pi G} \eta'_3\, ,
           \label{eq:1:12c}
\end{eqnarray}%
\end{mathletters}%
and,
\begin{mathletters}%
\label{eq:1:13all}%
\begin{eqnarray}
{\cal E}  &\equiv&  (D q)^a \epsilon_a{}^{b} p_b
           + \case{1}{2}  \left( \Pi^2 + {\varphi'}^2  \right)\, ,
           \label{eq:1:13a}  \\
{\cal P}  &\equiv&  - p_a (D q)^a - \Pi \varphi'\, ,
           \label{eq:1:13b}  %
\end{eqnarray}%
\end{mathletters}%
\begin{equation}
\hskip-10pt (D q)^a  \equiv  {q^a}' + \epsilon^a{}_{b}
           \left( q^b \omega_1 - e^b_1 \right)\, .\quad\quad {}
           \label{eq:1:14}
\end{equation}
(To conform to usual conventions,
some signs are changed from Ref.~\cite{ref:9}.)

The Poincar\'{e}-coordinate $q^a\/$ may be set to zero by a gauge
transformation, whereupon
(\ref{eq:1:11})--(\ref{eq:1:14}) can be shown to be equivalent
to~(\ref{eq:1:10}), as is demonstrated in the next subSection.
With dynamical $q^a\/$,
(\ref{eq:1:11})--(\ref{eq:1:14}) define a Poincar\'{e} gauge invariant
system --- that is why $q^a\/$ is like a Higgs field.  In our previous
investigation~\cite{ref:9}, we began with
(\ref{eq:1:11})--(\ref{eq:1:14}), which although not manifestly
gauge invariant, clearly exhibit that symmetry by virtue of the
symplectic structure in (\ref{eq:1:11}) and the vanishing of the gauge
constraints $(G_a, G_2, G_3)\/$.  In the next subSection, we show
explicitly how
the haphazard-looking expressions (\ref{eq:1:11})--(\ref{eq:1:14}) in
fact follow from a manifestly gauge invariant formalism.  The constraint
structure in the above equations is a guide to field redefinitions
and canonical transformations \cite{ref:9} that map the model onto
decoupled field variables entering quadratically in
(\ref{eq:1:15}) and (\ref{eq:1:16all}).

\subsection{Deriving the Constraint Structure}

In this subSection we present a gauge formulation of matter-coupled
dilaton gravity  (\ref{eq:1:4}), (\ref{eq:1:9}) and
give a derivation of Eqs.~(\ref{eq:1:11})--(\ref{eq:1:14}).
As mentioned before, the gauge structure
is based on a central extension of the two-dimensional Poincar\'e
algebra (\ref{eq:1:5all}) with four generators $\{Q_A\}_{A=0,1,2,3} =
\{P_a,J,I\}_{a=0,1}$. In the adjoint representation, a group element
parametrized by $\{ \theta^a,\alpha,\beta \}$ is a $4\times4$ matrix
\begin{equation}
U^A{}_B = \left(
  \begin{array}{ccc}
    \Lambda(\alpha)^a{}_b & - \epsilon^a{}_c \theta^c & 0 \\
    0 & 1 & 0 \\
    \theta^c \epsilon_{cd} \Lambda(\alpha)^d{}_b & -\case{1}{2} \theta^c
    \theta_c & 1
  \end{array}
\right)\, ,
\label{eq:2:1}
\end{equation}
with Lorentz transformation $\Lambda(\alpha)^a{}_b = \delta^a{}_b
\cosh\alpha + \epsilon^a{}_b \sinh\alpha$,
and no dependence on the parameter $\beta\/$.
A contravariant element $H = \eta^A
Q_A = \eta^a P_a + \eta^2 J + \eta^3 I\/$ transforms as $\eta^A \to
(U^{-1})^A{}_B \eta^B$, or
in components
\begin{mathletters}%
\label{eq:2:1.5all}%
\begin{eqnarray}
\eta^a  &\to&  \left( \Lambda^{-1} \right)^a {}_b
           \left( \eta^b + \eta^2 \epsilon^b {}_c \theta^c \right)\,,
           \label{eq:2:1.5a}  \\
\eta^2  &\to&  \eta^2\,,
           \label{eq:2:1.5b}  \\
\eta^3  &\to&  \eta^3 + \eta^a \epsilon_{a b} \theta^b
           - \case{1}{2} \eta^2 \theta^a \theta_a\,.
           \label{eq:2:1.5c}
\end{eqnarray}%
\end{mathletters}%
Notice that vectors along $I\/$ are invariant, a
property of a solvable algebra.

Since the extended Poincar\'e algebra is not semi-simple, it has a
degenerate Cartan-Killing form. There is, however,  a one-parameter
family of inner products
\begin{mathletters}%
\label{eq:2:2all}%
\begin{equation}%
 \langle Q_A \mid Q_B \rangle = h_{A B}\, ,
           \label{eq:2:2a}
\end{equation}
\begin{eqnarray}
h_{A B}  &=&  \left(
  \begin{array}{ccc}
    h_{a b} & 0 & 0 \\
    0 & c & -1 \\
    0 & -1 & 0
  \end{array}
 \right)\,,
    \label{eq:2:2b}  \\[1ex]
\noalign{\hbox{that are invariant
$\langle Q_C U^C{}_A \mid Q_D U^D{}_B \rangle = \langle Q_A \mid Q_B \rangle$,
and non-degenerate.  The inverse to $h_{A B}\/$ is}
\vskip1ex}
h^{A B}  &=&  \left(
  \begin{array}{ccc}
    h^{a b} & 0 & 0 \\
    0 & 0 & -1 \\
    0 & -1 & -c
  \end{array}
 \right)\,.
\label{eq:2:2c}
\end{eqnarray}%
\end{mathletters}%
It is convenient to set the parameter $c\/$ to zero. This is achieved
by shifting $J\/$ with a multiple of the central element~$I$, and we
make this choice henceforth.
The bilinear form defines a metric on the 4-dimensional (Lie algebra)
space, which is used to raise and lower indices, interchanging
covariant and contravariant tensors, {\it e.g.}~$\eta_A = h_{A B} \eta^B\/$
transforms as $\eta_A \to \eta_B U^B{}_A$, in components
\begin{mathletters}%
\label{eq:2:3all}%
\begin{eqnarray}%
\eta_a & \to &
           \left( \eta_b - \eta_3 \epsilon_{bc} \theta^c \right)
           \Lambda^b{}_a \,,
           \label{eq:2:3a}  \\
\eta_2 & \to &
           \eta_2 - \eta_a \epsilon^a{}_b \theta^b
           - \case{1}{2} \eta_3 \theta^a \theta_a \,,
           \label{eq:2:3b}  \\
\eta_3 & \to & \eta_3 \,.
           \label{eq:2:3c}
\end{eqnarray}
\end{mathletters}%

The gravitational fields are collected in the gauge potential
(\ref{eq:1:6a}) and the existence of the non-degenerate inner product
(\ref{eq:2:2all}) allows the construction of an action for pure dilaton gravity
\begin{equation}
I_{\rm gravity} = \int d^2x \, \Bigl\langle H \mid F \Bigr\rangle\,.
\label{eq:2:4}
\end{equation}
The solutions to the classical equations of motion,
\begin{equation}%
A_\mu = U^{-1} \partial_\mu U \,,\qquad
           \partial_\mu (U H U^{-1}) = 0\,,
\label{eq:2:5all}%
\end{equation}
are labeled by the two gauge invariants
$M = \langle H \mid H \rangle = \eta^a \eta_a - 2 \eta_2 \eta_3\/$
and $\lambda = \langle I \mid H \rangle = \eta_3$.
The norm square of $H\/$ corresponds to the ``mass'' $M\/$ of the CGHS
black-hole. Its projection on the invariant central element coincides
with the cosmological constant $\lambda\/$, which is not fixed but
enters as a dynamical variable in our model.

The coupling of matter to dilaton gravity in an explicit gauge invariant way
requires the use of two additional fields $q^a\/$ called
Poincar\'e coordinates \cite{ref:8a,ref:8}.  Consider a fiducial
quantity $q^A_{(0)}\/$ that is constant and picks out the Lorentz
generator~\cite{ref:14a}
\begin{equation}
q_{(0)} = q^A_{(0)} Q_A = J\, .
\label{eq:2:6}
\end{equation}
Using the Poincar\'e coordinates, we transform $q_{(0)}\/$ into the
$P_a\/$ direction and obtain a Lie algebra element that is taken to transform
contravariantly
\begin{equation}
q \equiv \exp \left( q^a \epsilon_a{}^b P_b \right)
           q_{(0)}  \exp
           \left( - q^a \epsilon_a{}^b P_b \right)
           = q^a P_a + J + \case{1}{2} q^a q_a I\,.
\label{eq:2:7}
\end{equation}
The transformation of this vector, $q^A \to (U^{-1})^A{}_B q^B\/$, induces a
transformation of the Poincar\'e coordinates, see~(\ref{eq:2:1.5a}),
\begin{equation}
q^a \to \left( \Lambda^{-1} \right)^a{}_b
           \left( q^b + \epsilon^b{}_c \theta^c \right)\, ,
\label{eq:2:8}
\end{equation}
which defines a non-linear action of the group~\cite{ref:8a}. Notice
that the Poincar\'e coordinates can always be set to zero with an
appropriate gauge transformation $\theta^b = - \epsilon^b{}_c q^c$.

Matter fields are described in a similar way.  Although we are
ultimately interested in massless fields, we find a manifestly
gauge invariant action only for massive scalar fields with mass $m\/$,
which we later set to zero.  Starting with the
combination
\begin{equation}
\Phi_{(0)} = \frac{\varphi^a}{m^2} P_a - \varphi_3 J\, ,
\end{equation}
we construct
\begin{equation}
\Phi \equiv \exp \left( q^a \epsilon_a{}^b P_b \right)   \Phi_{(0)}
           \exp \left( - q^a \epsilon_a{}^b P_b \right)
           = \left( \frac{\varphi^a}{m^2}
                      - \varphi_3 q^a \right)  P_a - \varphi_3 J
           + \left( \frac{\varphi^a}{m^2}
                      - \case{1}{2} \varphi_3 q^a  \right)  q_a I\,,
\label{eq:2:10}
\end{equation}
and choose it to transform contravariantly under a gauge
transformation.  According to (\ref{eq:2:1.5all}),
this induces the matter field transformation law
\begin{mathletters}%
\label{eq:2:11all}%
\begin{eqnarray}%
 \varphi^a &\to& (\Lambda^{-1})^a{}_b \varphi^b\,,
           \label{eq:2:11a}  \\
 \varphi_3 &\to& \varphi_3\,.
           \label{eq:2:11b}
\end{eqnarray}
\end{mathletters}%
Gravity is minimally coupled through the covariant derivative,
and the matter-gravity interaction is taken in explicitly gauge
invariant form
\begin{equation}%
I_{\rm matter} = \frac{m^2}{4} \int d^2x
           \left\langle q  \Biggl| \epsilon^{\mu\nu} [D_\mu \Phi, D_\nu \Phi]
           + \frac{m^2}{2} e (q) \Bigl[ \Phi, [\Phi, q] \Bigr]
           \right\rangle\,,
\label{eq:2:12}
\end{equation}
where
\begin{eqnarray}%
D_\mu  &=&  \partial_\mu + [A_\mu, \;\; ]\,,
           \nonumber \\
e (q)  &=&  \case{1}{2} \epsilon^{\mu \nu}
                      \epsilon_{a b} (D_\mu q)^a (D_\nu q)^b\,.
           \label{eq:2:12.5}
\end{eqnarray}
Together with the dilaton gravity action (\ref{eq:2:4}), (\ref{eq:2:12})
describes two-dimensional matter-coupled dilaton gravity
in a manifestly topological and gauge invariant way.

The relation to the Klein-Gordon action becomes apparent if we rewrite
(\ref{eq:2:12}) using the components of the fields. Up to surface
terms, which we drop, $I_{\rm matter}\/$ is
\begin{eqnarray}
I_{\rm matter} &=& - \case{1}{2} \int d^2 x
           \left[ \varphi_3 \varphi_a f^a + \varphi_a
           \biggl (\frac{1}{2} \frac{\varphi^a}{m^2} - \varphi_3 q^a \biggr)
           f^2 \right]
           \nonumber \\
&&{\quad} + \case{1}{8} \int d^2x \, e(q)
           \left( \varphi^a + 2 \epsilon^{a b} E(q)^\mu_b \,
           \partial_\mu \varphi_3 \right)
           \Bigl( \varphi_a + 2 \epsilon_a{}^c E(q)^\nu_c \,
           \partial_\nu \varphi_3 \Bigr)
           \nonumber\\
&&{\quad} + \case{1}{2} \int d^2x \, e(q)
           \left[ h^{a b} \, E(q)^\mu_a \, E(q)^\nu_b \,
           \partial_\mu \varphi_3 \, \partial_\nu \varphi_3
           - m^2 \varphi_3{}^2 \right]\,,
           \label{eq:2:13} \\
\noalign{\hbox{with}}
E(q)^\mu_a  &=&  \frac{1}{e(q)} \epsilon^{\mu\nu} h_{a b}
           (D_\nu q)^b\, .
           \label{eq:2:13a}
\end{eqnarray}
The first term in the right side of (\ref{eq:2:13}) is absorbed in a
redefinition
of $\eta_a\/$ and $\eta_2\/$ in the pure dilaton gravity
Lagrangian~(\ref{eq:1:8}), (\ref{eq:2:4}),
whereupon the massless limit can be taken.
The second term sets
$\varphi^a\/$ equal to the gradient of the field $\varphi_3\/$. In the
gauge $q^a = 0$, $e(q)\/$ becomes the density
$\sqrt{-g}\/$, $E(q)^\mu_a\/$ the inverse {\it Zweibein} $E^\mu_a$
from which the metric is reconstructed: $h^{a b} \, E^\mu_a \, E^\nu_b
= g^{\mu\nu}$, and the third term coincides with the action of a
massive scalar field~\cite{ref:14b}.

Quantization is carried out in a first order formalism and we want to
show now that (\ref{eq:1:11})--(\ref{eq:1:14}) follow from
(\ref{eq:2:4}) and~(\ref{eq:2:13}).  As said before, the first term on
the right side of (\ref{eq:2:13}) is combined with the dilaton gravity action
by a redefinition of $\eta_a\/$ and $\eta_2\/$.  The spatial
components of the gauge fields multiply $\dot{\eta}_A\/$ and become
the momenta conjugated to $\eta_A$, whereas the gauge field time
components play the role of Lagrange multipliers.  In the remainder of
the first order expression for~(\ref{eq:2:13}), the coefficients of
$\dot{q}^a\/$ and $\dot{\varphi}_3\/$ are identified with their
respective canonical momenta $p_a\/$ and $\Pi\/$, whose definition is
enforced with three Lagrange multipliers $\{ N^a,N \}\/$.
Consequently, ${\cal L}_{g + m} = \frac{1}{4 \pi G} {\cal L}_{\rm
gravity} + {\cal L}_{\rm matter}\/$ is written in an equivalent form
as
\begin{eqnarray}%
{\cal L}_{g + m} &=& \frac{1}{4\pi G}
           \left( \eta_a \dot e^a_1 + \eta_2 \dot\omega_1
		+ \eta_3 \dot a_1 \right)
           + p_a \dot{q}^a + \pi \dot{\varphi}_3
           + e^a_0 G_a + \omega_0 G_2
           + a_0 G_3
           \nonumber \\
&& {\quad} + N^a \left[  p_a - \case{1}{2} \epsilon_{a b} \varphi^b \varphi'_3
           - \case{1}{8} \epsilon_{a b} (D q)^b
           \left( \varphi^c \varphi_c - 4 m^2 \varphi^2_3 \right) \right]
           \nonumber \\
&& {\quad} + N \left[ \Pi - \case{1}{2}
                      \varphi^a \epsilon_{a b} (D q)^b \right]\,,
           \label{eq:2:14}
\end{eqnarray}
where the $\{ G_A \}_{A=0,1,2,3}\/$ are given by (\ref{eq:1:12all}) and
$D\/$ denotes the spatial component of the covariant derivative
(\ref{eq:2:12.5}).  The three constraints
enforced by $\{ N^a, N \}\/$ may be replaced by three alternative
ones, enforced by $\{ u,v,w\}$,
\begin{eqnarray}
&-&  u \left[ - p_a \epsilon^a{}_b (D q)^b  + \case{1}{2}
           \left\{\Pi^2 + \varphi'_3{}^2
           - \left[ \varphi_3' + \case{1}{2} \varphi_a (D q)^a \right]^2
           + m^2 (D q)^a (D q)_a \varphi_3{}^2
           \right\} \right]
           \nonumber \\
&-& v \Bigl[ - p_a (D q)^a - \Pi \varphi_3' \Bigr]
           \nonumber \\
&-&  w \left[p_a (D q)^a + \case{1}{2} \varphi^a
           \epsilon_{a b} (D q)^b \varphi'_3 \right]\,.
           \nonumber
\end{eqnarray}
Since no derivative acts on the $\varphi^a$, we eliminate them through their
equations of motion
\begin{mathletters}%
\label{eq:2:15all}
\begin{equation}%
u \, (D q)^a \left( \varphi_3' + \case{1}{2} \varphi_b (D q)^b \right)
           - w \, \epsilon^a{}_b (D q)^b \varphi'_3 = 0\,.
\label{eq:2:15a}
\end{equation}
Successive projections of Eq.~(\ref{eq:2:15a}) onto $(D q)_a$ and
$\epsilon_{a b} (D q)^b$ imply
\begin{equation}
w = 0
\,,\qquad
\varphi_3' + \case{1}{2} \varphi_a (D q)^a = 0\,.
\label{eq:2:15b}
\end{equation}%
\end{mathletters}%
In the massless case, (\ref{eq:2:14}) and (\ref{eq:2:15b})
reproduce (\ref{eq:1:11})--(\ref{eq:1:14}), once
$\varphi_3\/$ is identified with $\varphi\/$.

\section{Metric-Based Starting Point}
\label{sec:3}

Beginning with the metric formulation of the action (\ref{eq:1:10}) as
in (\ref{eq:1:4}) and~(\ref{eq:1:9}), we derive (\ref{eq:1:15}) and
(\ref{eq:1:16all}), without using gauge theoretical ideas.  To begin,
make the following parameterization for the metric tensor~\cite{ref:6}.
\begin{equation}
g_{\mu\nu}
           = e^{2\rho}
           \pmatrix{u^2-v^2 & v  \cr
                      v &-1 \cr }\,.
\label{eq:3:1}
\end{equation}
Passing to canonical form, the action (\ref{eq:1:10}) becomes
\begin{equation}
I = \int d^2 x
           \Bigl[ \Pi_\rho \dot \rho
           + \Pi_\eta \dot \eta + \Pi \dot \varphi
           - u {\cal E} - v {\cal P}
           \Bigr]\,,
\label{eq:3:2}
\end{equation}
where the constraints are
\begin{mathletters}%
\label{eq:3:3all}
\begin{eqnarray}
{\cal E} &=&  - \frac{1}{2 \pi G} (\eta'' - \rho' \eta')
           +2 \pi {G} \Pi_\rho \Pi_\eta
           + {\lambda \over 4 \pi {G}} e^{2\rho}
           + \case{1}{2} (\Pi^2 + \varphi'^2)\,,
           \label{eq:3:3a}  \\
{\cal P} &=& - \Pi_\rho \rho'
           + \Pi_\rho' - \Pi_\eta \eta'
           - \Pi \varphi'\,.
           \label{eq:3:3b}
\end{eqnarray}%
\end{mathletters}%

The variables $\rho$, $\Pi_\rho$, $\eta$ and $\Pi_\eta$ are changed to
new variables $\rho^a$ and $p_a (a=0,1)$ using the canonical
transformation induced by the generating functional
\begin{eqnarray}
F(\rho, \Pi_\eta\, ; \rho^0,\rho^1) &=& -
\int_{-\infty}^{\infty} d\sigma \, e^{\rho(\sigma)} ~
\left[
\rho^0 (\sigma) \sinh \theta(\sigma)
- \rho^1 (\sigma) \cosh \theta(\sigma)
\right]\,,
\label{eq:3:4} \\
\theta(\sigma) &\equiv& 2 \pi G
\int_{-\infty}^{\sigma} d \tilde{\sigma} \, \Pi_\eta(\tilde{\sigma})\,.
\label{eq:3:5}
\end{eqnarray}
(All of the fields in this expression have the same time argument, which
is suppressed; $\sigma$ and $\tilde{\sigma}$ are spatial variables).
The new variables $\rho^a(\sigma)$ and $p_a(\sigma)$ obey
$p_a = {-\delta F / \delta \rho^a}$, while similarly
for the old variables, $\eta = - \delta F / \delta {\Pi_\eta}$
and $\Pi_\rho = \delta F / \delta \rho$. Note that the above canonical
transformations are non-local in the spatial variable $\sigma$
but local in time.

In these variables the constraints
(\ref{eq:3:3all}) become
\begin{mathletters}%
\label{eq:3:11all}
\begin{eqnarray}
{\cal E}  &=&  - \rho^{a \prime}\epsilon_a{}^{b} p_b
           - {\lambda \over 4 \pi {G}} p_a p^a
           + \case{1}{2}  \left( \Pi^2 + \varphi'^{2} \right)\,,
           \label{eq:3:11a}  \\
{\cal P}  &=&  - p_a \rho^{a \prime}
           - \Pi \varphi'\,.
           \label{eq:3:11b}
\end{eqnarray}%
\end{mathletters}%
This form of the constraints appears at an intermediate step in~\cite{ref:9}.
In order to make the comparison with (\ref{eq:1:16all})
complete, we make the transformation
\begin{mathletters}%
\label{eq:3:12all}
\begin{eqnarray}
\pi_a  &=&  {\lambda \over 4 \pi {G}} p_a
           - \case{1}{2} \epsilon_{ab} \rho^{b\prime}\,,
           \label{eq:3:12a}  \\
r^a  &=&  {4 \pi {G} \over \lambda} \rho^a \,,
           \label{eq:3:12b}
\end{eqnarray}%
\end{mathletters}%
and find that (\ref{eq:3:11all}) coincide
with~(\ref{eq:1:16all})~\cite{ref:10}.

\section{Removing Anomalies in the Canonical Theory}
\label{sec:4}

The theory described by (\ref{eq:1:15}) and (\ref{eq:1:16all}) appears
to be very simple: there are three independent fields
$\{ r^a, \varphi \}\/$ and together with the canonical momenta
$\{ \pi_a, \Pi \}\/$ they lead to a quadratic Hamiltonian, which has no
interaction terms among the three; see~(\ref{eq:1:16a}).  Similarly,
the momentum comprises non-interacting terms, see~(\ref{eq:1:16b}).
However, there remains a subtle ``correlation interaction'' as a
consequence of the requirement that the energy and momentum densities,
$\cal E\/$ and $\cal P\/$, annihilate physical states, as follows from
varying the Lagrange multipliers $u\/$ and $v\/$ in~(\ref{eq:1:15})
\begin{mathletters}%
\label{eq:4:1all}%
\begin{eqnarray}
{\cal E} \mid \psi \rangle &=& 0\,,
           \label{eq:4:1a}  \\
{\cal P} \mid \psi \rangle &=& 0\,.
           \label{eq:4:1b}
\end{eqnarray}%
\end{mathletters}%
Thus, even though $\cal E\/$ and $\cal P\/$ each are sums of
non-interacting variables, the physical states $\mid \psi \rangle\/$
are not direct products of states for the separate degrees of freedom.
Note that Eqs.~(\ref{eq:4:1all}) comprise the entire physical content
of the theory.  There is no need for any further ``gauge fixing'' or
``ghost'' variables --- this is the advantage of the Hamiltonian
formalism that we are pursuing in this Section.

The momentum constraint (\ref{eq:4:1b}) is easy to unravel: in a
Schr\"{o}dinger representation where $\mid \psi \rangle\/$ is realized
as a functional $\Psi\/$ of $\{ r^a, \varphi \}\/$, on which these
quantities act by multiplication while the canonically conjugate
momenta $\{ \pi_a, \Pi \}\/$ act by (functional) differentiation,
Eq.~(\ref{eq:4:1b}) implies that $\Psi\/$ is a functional that is
invariant against arbitrary reparameterization of the spatial
$\sigma$-variable: $\sigma \to \tilde{\sigma} (\sigma)$.
Such functionals are readily
constructed.  Moreover, the commutator of $\cal P\/$ with itself
closes on itself --- even in the quantum theory the constraint is
first class and there is no obstruction to~(\ref{eq:4:1b})
\begin{equation}
i \Bigl[ {\cal P} (\sigma), {\cal P} (\tilde{\sigma}) \Bigr]
           = \Bigl( {\cal P} (\sigma) + {\cal P} (\tilde{\sigma}) \Bigr)
           \delta' (\sigma - \tilde{\sigma})\,.
\label{eq:4:2}
\end{equation}
(The commutator is at equal times,
and the time argument is suppressed.)

The constraint (\ref{eq:4:1a}) is the Wheeler-DeWitt equation, and
here an anomaly obstructs solving it: while the commutator of
$\cal E\/$ with itself closes on $\cal P\/$~\cite{ref:14c}
\begin{equation}
i \Bigl[ {\cal E} (\sigma), {\cal E} (\tilde{\sigma}) \Bigr]
           = \Bigl( {\cal P} (\sigma) + {\cal P} (\tilde{\sigma}) \Bigr)
           \delta' (\sigma - \tilde{\sigma})\,,
\label{eq:4:3}
\end{equation}
the $[ \cal E, P ]\/$ commutator possesses, in addition to an
innocuous term involving $\cal E\/$, a quantum $c\/$-number anomaly
\begin{equation}
i \Bigl[ {\cal E} (\sigma), {\cal P} (\tilde{\sigma}) \Bigr]
           = \Bigl( {\cal E} (\sigma) + {\cal E} (\tilde{\sigma}) \Bigr)
           \delta' (\sigma - \tilde{\sigma})
           - \frac{1}{12 \pi} \delta''' (\sigma - \tilde{\sigma})\,.
\label{eq:4:4}
\end{equation}
The triple-derivative Schwinger term converts the classical first
class constraint into a quantal second class one and prevents finding
a solution to~(\ref{eq:4:1a}).  Note that the Schwinger term arises
solely from the matter terms; the Schwinger term in the gravitational
variables vanishes owing to the indefinite sign: the contribution from
$a = 0\/$ cancels against that from $a = 1\/$.  Correspondingly, in
the absence of matter variables, all constraints can be, and have been
solved --- there is no obstruction in the gravity
sector.

In the absence of matter one finds two gravity states
$| \pm \rangle_{\rm gravity}$. They
are explicitly represented by the functionals
\begin{equation}
\Psi_{\rm gravity} (r^a)
= \exp \Bigl( {\pm i {\Lambda\over2} \int d\sigma
\epsilon_{ab} r^a  {r^b}'}\Bigr)\,,
\label{eq:4:5new}
\end{equation}
and satisfy the gravitational portions of the constraints
(\ref{eq:1:16all}) \cite{ref:9}
\begin{mathletters}%
\label{eq:4:6allnew}%
\begin{eqnarray}
-{1\over2}
\left( {1\over\Lambda}\pi^a \pi_a + \Lambda {r^a}' r'_a\right)\Psi_{\rm
gravity}
&=& {1\over2} \left({1\over\Lambda}
{\delta^2 \over \delta r^a \delta r_a} - \Lambda {r^a}' r'_a\right)
\Psi_{\rm gravity} = 0\,,
\label{eq:4:6anew} \\
(-{r^a}' \pi_a) \Psi_{\rm gravity}
&=&
\left( i \, {r^a}' {\delta \over \delta r^a} \right)\Psi_{\rm gravity} = 0\,.
\label{eq:4:6bnew}
\end{eqnarray}%
\end{mathletters}%
The states (\ref{eq:4:5new}) annihilated by the constraints may also be
related to the Fock vacuum $| 0 \rangle$, which in the Schr\"odinger
representation reads
\begin{eqnarray}%
\Psi_{\rm vacuum} (r^a)
&=& \det{}^{\!1\over2} \left( {\Lambda\omega \over \pi} \right) \cdot
\exp \Bigl( {-{\Lambda\over2} \int d{\sigma} d{\tilde\sigma}
(r^0 \omega r^0 + r^1 \omega r^1)} \Bigr) \,,
\nonumber\\
\omega(\sigma, \tilde{\sigma}) &=&
\int {dk \over 2\pi}
e^{-ik(\sigma - \tilde{\sigma})} |k|
\,.
\label{eq:4:7new}
\end{eqnarray}
The gravity states are then
given by
\begin{equation}
| \pm \rangle_{\rm gravity} = \det{}^{-{1\over 2}}\left(
{\Lambda\omega \over 2\pi} \right) \cdot
\Bigl[ \exp \Bigl( {\pm {\Lambda \over 2} \int {dk}
\, a_0^{\dagger}(k) \, \epsilon(k) \, a_1^{\dagger} (-k)} \Bigr) \Bigr]
| 0 \rangle\,,
\label{eq:4:8new}
\end{equation}
where the creation operators are defined in the same way for both fields
\beq
a_a^{~\dagger} (k) =
{-i \over \sqrt{4\pi|\Lambda k|}}
\int d\sigma \, e^{ik\sigma} \, \pi_a(\sigma)
{~+~} \sqrt{|\Lambda k|\over4\pi}
\int d\sigma \, e^{ik\sigma} \, r^a(\sigma)\,.
\label{eq:4:9new}
\eeq

These pure dilaton gravity states, satisfying all constraints,
have been obtained
within the gauge theoretical formalism in Refs. \cite{ref:9,ref:5},
within the metric formalism in Ref. \cite{ref:6}, and the relation
between the two approaches has been elucidated in Ref.\cite{ref:7}.

Once matter degrees of freedom are included, however,
our Hamiltonian analysis of the theory cannot be taken farther
owing to the Schwinger term.
We now show that it is possible to alter the theory, by a change in
the gravitational sector, whereupon the anomaly cancels.

To describe this modification, it is first convenient to form the sum
and difference of the constraints, putting them into the decoupled,
Virasoro form,
\begin{eqnarray}
\Theta_{\pm}  &=& \case{1}{2} ({\cal E \mp P})\,,
           \label{eq:4:5} \\
{\cal L}  &=& \pi_a \dot{r}^a + \Pi \dot{\varphi}
           - \lambda^{+} \Theta_{+} - \lambda^{-} \Theta_{-}\,,
           \nonumber \\
\lambda^{\pm}  &=& u \mp v\,,
           \label{eq:4:6}
\end{eqnarray}
with (\ref{eq:4:2})--(\ref{eq:4:4}) becoming
\begin{mathletters}%
\label{eq:4:7all}%
\begin{eqnarray}
\Bigl[ \Theta_{\pm} (\sigma) , \Theta_{\pm} (\tilde{\sigma}) \Bigr]
           &=&   \pm i \Bigl[ \Theta_{\pm} (\sigma)
           + \Theta_{\pm} (\tilde{\sigma}) \Bigr]
           \delta' (\sigma - \tilde{\sigma})
           \mp \frac{i}{24 \pi} \delta''' (\sigma - \tilde{\sigma})\,,
           \label{eq:4:7a}  \\  {}
\Bigl[ \Theta_{\pm} (\sigma) , \Theta_{\mp} (\tilde{\sigma}) \Bigr]
           &=&  0 \,.
           \label{eq:4:7b}
\end{eqnarray}%
\end{mathletters}%
Next, following Kucha\v{r}~\cite{ref:12}, we pass to new canonical
variables, with the transformation
\begin{mathletters}%
\label{eq:4:8all}%
\begin{eqnarray}
P_{\pm}  &=&  - \frac{1}{2 \sqrt{\Lambda}} (\pi_0 + \pi_1)
           \pm \frac{\sqrt{\Lambda}}{2}
           \left( r^{0 \prime} - r^{1 \prime} \right)\,,
           \label{eq:4:8a}  \\
X^{\pm \prime}  &=&  \pm \frac{1}{2 \sqrt{\Lambda}} (\pi_0 - \pi_1)
           - \frac{\sqrt{\Lambda}}{2}
           \left( r^{0 \prime} + r^{1 \prime} \right)\,,
           \label{eq:4:8b}
\end{eqnarray}%
\end{mathletters}%
in terms of which the constraints read
\begin{equation}
\Theta_{\pm}  =  \pm P_{\pm} X^{\pm \prime} + \theta_{\pm}\,,
\label{eq:4:9}
\end{equation}
where $\theta_{\pm}\/$ is just the matter part
\begin{equation}
\theta_{\pm}  =  \case{1}{4} (\Pi \pm \varphi')^2\,.
\label{eq:4:10}
\end{equation}

The gravitational contribution to $\Theta_{\pm}\/$ has been
transformed into $\pm P_{\pm} X^{\pm \prime}\/$, which looks like the
momentum density for fields $\{ P_{\pm}, X^{\pm} \}\/$.
One sees once again that the gravity portions of the constraint do not
give rise to an anomaly: a momentum density commutator possesses no
anomaly, see~(\ref{eq:4:2}) --- it is present only in the matter-part
commutator $[ \theta_{\pm}, \theta_{\pm} ]\/$.

With Kucha\v{r}~\cite{ref:12}, we now ask: is it possible to change
the theory by adding something to $\Theta_{\pm}\/$, so that the
modified constraints possess no anomaly?  A remarkably simple
expression is found to do the job.  One verifies that the algebra of
$\widetilde{\Theta}_{\pm}\/$, defined by
\begin{eqnarray}
\widetilde{\Theta}_{\pm}  &=&  \Theta_{\pm} + \frac{1}{48 \pi}
           (\ln X^{\pm \prime})''\,,
           \nonumber  \\
&=&  \pm P_{\pm} X^{\pm \prime} + \frac{1}{48 \pi}
           \left( \frac{X^{\pm \prime\prime\prime}}{X^{\pm \prime}}
           - \left( \frac{X^{\pm \prime\prime}}{X^{\pm \prime}} \right)^2
           \right)
           + :\hskip-1pt\theta_{\pm}\hskip-1pt :\,\,,
\label{eq:4:11}
\end{eqnarray}
possesses no anomaly.  Here, $:\hskip-1pt\theta_{\pm}\hskip-1pt :$
is normal-ordered
with respect to the Fock vacuum defined in (\ref{eq:4:7new}).

[Actually Kucha\v{r}'s approach \cite{ref:12} is different: he normal orders
$\theta_{\pm}\/$ with respect to a Gaussian (Fock) vacuum with covariance
depending on $X^{\pm}\/$; this changes the anomaly in the matter
commutator $[ \theta_{\pm}, \theta_{\pm} ]\/$ to an
$X^{\pm}\/$-dependent expression.  He then adds a further
$X^{\pm}\/$-dependent term to $\Theta_{\pm}\/$, whose effect is to
cancel the modified anomaly.  In our approach, we remain with the
conventionally ordered $\theta_{\pm}\/$ and the conventional anomaly,
and find that the relatively simple addition in~(\ref{eq:4:11}) is
sufficient.  Moreover, (\ref{eq:4:11}) has a natural interpretation,
see below.]

With $\widetilde{\Theta}_{\pm}\/$ there is no obstruction, and we are
instructed to solve
\begin{mathletters}%
\label{eq:4:12all}%
\begin{equation}
\widetilde{\Theta}_{\pm} \mid \psi \rangle = 0\,,
\label{eq:4:12a}
\end{equation}%
or equivalently, in the Schr\"{o}dinger representation
\begin{equation}
\left( \frac{1}{i}  \frac{\delta}{\delta X^{\pm}}
           \pm \frac{1}{48 \pi X^{\pm \prime}}
           (\ln X^{\pm \prime})''
           \pm \frac{1}{X^{\pm \prime}} \theta_{\pm} \right)
           \Psi = 0\,.
\label{eq:4:12b}
\end{equation}%
\end{mathletters}%
One may say \cite{ref:12,ref:14}
that the anomaly is removed by introducing functional
U(1) connections
\begin{equation}%
{\cal A}_{\pm} (X^{\pm}) = \pm \frac{1}{48 \pi X^{\pm \prime}}
           (\ln X^{\pm \prime})'' \,,
\label{eq:4:13a}
\end{equation}
with curvature $\delta {\cal A}_{\pm}
\left( X^{\pm} (\tilde{\sigma}) \right)/ \delta X^{\pm}(\sigma)
- \delta {\cal A}_{\pm}
\left( X^{\pm} (\sigma) \right) /\delta X^{\pm}(\tilde{\sigma})\/$.
In the modified constraint $\widetilde{\Theta}_{\pm}\/$, one still
observes that there is no mixing between gravitational variables
$\{ P_{\pm}, X^{\pm} \}\/$ and matter variables $\{ \Pi, \varphi\}\/$.
But the modified gravitational contribution is no longer quadratic ---
indeed it is non-polynomial --- and we have no idea how to
solve~(\ref{eq:4:12all})~\cite{ref:15a}.

The addition that we have made can be related to structures that
have already appeared in the literature in descriptions of
two-dimensional matter fields interacting with external, c-number
gravity~\cite{ref:13}. We now explain this.

Observe that our
modified action reads
\begin{eqnarray}
\widetilde{I}  &=&  \int d^2 x \Biggl\{
           P_{+} \dot{X}^{+} + P_{-} \dot{X}^{-} + \Pi \dot{\varphi}
           \nonumber \\
&& {\quad} - \lambda^{+} \left(P_{+} X^{+ \prime} + \frac{1}{48 \pi}
           (\ln X^{+ \prime})'' + \theta_{+} \right)
           \nonumber \\
&& {\quad} - \lambda^{-} \left( - P_{-} X^{- \prime} + \frac{1}{48 \pi}
           (\ln X^{- \prime})'' + \theta_{-} \right)
\Biggr\}\,.
\label{eq:4:14}
\end{eqnarray}
Eliminating $P_{\pm}\/$ evaluates $\lambda^{\pm}\/$ as
$\pm \dot{X}^{\pm} / X^{\pm \prime}\/$ and leaves~\cite{ref:15}
\begin{eqnarray}
\widetilde{I}  &=&  I_0 + \Delta I\,,
           \label{eq:4:15}  \\
I_0  &=&  \int d^2 x
           \left( \Pi \dot{\varphi}
                      - \frac{\dot{X}^{+}}{X^{+ \prime}} \theta_{+}
                     + \frac{\dot{X}^{-}}{X^{- \prime}} \theta_{-} \right)\,,
           \label{eq:4:16}  \\
\Delta I   &=&  \frac{1}{48 \pi} \int d^2 x
           \left(- \frac{\dot{X}^{+}}{X^{+ \prime}} (\ln X^{+ \prime})''
           + \frac{\dot{X}^{-}}{X^{- \prime}}
           (\ln X^{- \prime})''   \right)\,.
           \label{eq:4:17}
\end{eqnarray}
$I_0\/$ is recognized to be just the matter action~(\ref{eq:1:9}),
written in first order from (so that metric tensor components enter
only as Lagrange multipliers) with $g_{\mu \nu}\/$ parametrized as
\begin{eqnarray}
g_{\mu \nu}  &=&  e^{\chi} \partial_{\mu} X^{a} \partial_{\nu} X^{b} h_{a b}\,,
           \label{eq:4:18}  \\
\noalign{\hbox{and}}
\theta_{\pm}  &=&  \case{1}{4}
           \left( \sqrt{-g} \, g^{0 \mu} \partial_\mu \varphi \pm \varphi'
           \right)^2
           = \case{1}{4} \left( \Pi \pm \varphi' \right)^2\,.
           \label{eq:4:18.5}
\end{eqnarray}

It is well known that the energy momentum tensor $\theta_{\mu \nu}\/$
for the matter variables of the theory described by $I_0\/$
\begin{equation}
\theta_{\mu \nu} = \frac{2}{\sqrt{-g}} \frac{\delta I_0}{\delta g^{\mu \nu}}\,,
\label{eq:4:19}
\end{equation}
exhibits a quantal anomaly, which may be viewed as a diffeomorphism
anomaly or a trace ano\-maly.  Usually this is described by computing
the matrix elements of $\theta_{\mu \nu}\/$ in the presence of a
background $g_{\mu \nu}\/$, $\langle \theta_{\mu \nu} \rangle_g,\/$
and examining the divergence and trace of $\langle \theta_{\mu \nu}
\rangle_g\/$~\cite{ref:16}.  Alternatively one may functionally
integrate the matter variables, obtain a nonlocal effective
gravitational action, $I_{\rm effective} (g)\/$, which is functional
of $g_{\mu \nu}\/$, and study its diffeomorphism and Weyl invariance
properties~\cite{ref:2}.  Evidently $\langle \theta_{\mu \nu}
\rangle_g = (2 / \sqrt{-g} )
(\delta I_{\rm effective} (g) / \delta g^{\mu \nu})\/$.

Still another approach is the following~\cite{ref:13}.  Form
$\theta_{\mu \nu}\/$ as in (\ref{eq:4:19}) and express it in terms of
$\theta_{+}\/$ and $\theta_{-}\/$ in (\ref{eq:4:18.5}), viewing these
to be quantum operators satisfying the commutator
algebra~(\ref{eq:4:7all})
\begin{mathletters}%
\label{eq:4:20all}%
\begin{eqnarray}
\theta_{++}   &=&   \case{1}{2} (\lambda^{+} + 1)^2 \theta_{+}
           + \case{1}{2} (\lambda^{-} - 1)^2 \theta_{-}\,,
           \label{eq:4:20a} \\
\theta_{--}   &=&   \case{1}{2} (\lambda^{+} - 1)^2 \theta_{+}
           + \case{1}{2} (\lambda^{-} + 1)^2 \theta_{-}\,,
           \label{eq:4:20b}  \\
\theta_{+-}   &=&  \case{1}{2} \left( (\lambda^{+})^2  - 1 \right) \theta_{+}
           + \case{1}{2} \left( (\lambda^{-})^2  -1 \right) \theta_{-}\,\,.
           \label{eq:4:20c}
\end{eqnarray}%
\end{mathletters}%
Here $\lambda^{\pm}\/$ is the c-numbers $\lambda^{\pm} =
 (- \sqrt{-g} \, \pm g_{0 1})/g_{11}\/$.  It follows that the
operator $\theta_{\mu \nu}\/$ is traceless.  Next compute the
covariant divergence of $\theta_{\mu \nu}\/$, where time derivatives
are calculated as commutators with the Hamiltonian,
\begin{eqnarray}
H  &=&  \int d \sigma ( \lambda^{+} \theta_{+} + \lambda^{-} \theta_{-} )\,,
           \label{eq:4:21}  \\
\dot{\theta}_{\pm}  &=&  i [ H, \theta_{\pm} ]\,,
           \label{eq:4:22}
\end{eqnarray}
and the quantum anomaly in the $[ \theta_{\pm}, \theta_{\pm} ]\/$
commutator is taken into account.  One finds non-vanishing c-numbers
for the divergence
\begin{mathletters}%
\label{eq:4:23all}%
\begin{eqnarray}
D_{\mu} \theta^{\mu}{}_{ +}   &=&  - \frac{1}{24 \pi \sqrt{-g}}
           \frac{1}{\sqrt{2}}
           \left( ( \lambda^{+} + 1) \lambda^{+ \prime\prime\prime}
           - (\lambda^{-} - 1) \lambda^{- \prime\prime\prime} \right)\,,
           \label{eq:4:23a} \\
D_{\mu} \theta^{\mu}{}_{ -}  &=&  - \frac{1}{24 \pi \sqrt{-g}}
           \frac{1}{\sqrt{2}}
           \left( ( \lambda^{+} - 1) \lambda^{+ \prime\prime\prime}
           - ( \lambda^{-} + 1) \lambda^{- \prime\prime\prime} \right)\,.
           \label{eq:4:23b}
\end{eqnarray}%
\end{mathletters}%
Finally one asks whether there is a counter action $\Delta I\/$,
constructed solely from $g_{\mu \nu}\/$, which when summed with
$I_0\/$
\begin{equation}
\widetilde{I} = I_0 + \Delta I\,,
\label{eq:4:24}
\end{equation}
produces a traceless and conserved energy-momentum tensor
\begin{mathletters}%
\label{eq:4:25all}%
\begin{eqnarray}
\widetilde{\theta}_{\mu \nu}
           &=& \frac{2}{\sqrt{-g}}
                      \frac{\delta \widetilde{I}}{\delta g^{\mu \nu}}\,,
           \label{eq:4:25a} \\
D_\mu \widetilde{\theta}^\mu{}_\nu  &=&  0\,.
           \label{eq:4:25b}
\end{eqnarray}%
\end{mathletters}%
The answer is that $\Delta I\/$ in (\ref{eq:4:17}) does the job, with
$\lambda^{\pm} = \pm \dot{X}^{\pm} / X^{\pm \prime}\/$,
because the c-number covariant divergence of
$(2/\sqrt{-g}) \delta \Delta I/\delta g^{\mu \nu}\/$
cancels the right sides of (\ref{eq:4:23all}).

We conclude this Section with various observations.

\begin{enumerate}

\item
Even though $I_0\/$ in (\ref{eq:4:16}) is a matter action
like~(\ref{eq:1:9}), but presented in first-order form, the dynamical
equations that follow from (\ref{eq:4:16}) are not identical to those
of~(\ref{eq:1:9}).  As was seen, our transformations on the CGHS
theory result in the parameterization (\ref{eq:4:18}) for the
effective metric
tensor, which occurs in the final matter action~(\ref{eq:4:16}).  When
the metric tensor is varied as a whole, which one can do in
(\ref{eq:1:9}), one gets the
(classical)
equation that $\theta_{\mu \nu}\/$ must
vanish.  On the other hand, when $g_{\mu \nu}\/$ is given
by~(\ref{eq:4:18}), one can only vary $\chi\/$ and $X^a\/$; this
gives weaker conditions: $\theta^\mu{}_\mu = 0\/$ and
$\theta^{\mu \nu} (D_\mu V^a_\nu + D_\nu V^a_\mu - g_{\mu \nu} D^{\alpha}
V^a_{\alpha}) =0\/$, where
$V^a_\mu = (\exp {\chi}) \partial_{\mu} X^a\/$
and (classical) conservation of $\theta^{\mu \nu}\/$ has been used.

\item
The consistent theory that we have constructed is diffeomorphism
invariant because the commutator anomaly has been removed, and also it is
Weyl invariant because the compensating term $\Delta I\/$ does not involve
the conformal factor of the metric.  This is possible because the
counter term, which is a local expression in terms of $X^{\pm}\/$ that
are the natural variables of the (transformed) CGHS theory, would be
nonlocal if expressed in terms of a metric tensor.

\item
There is another route to the conclusion that elimination of the
gravitational variables in~(\ref{eq:1:15}), (\ref{eq:1:16all}) results
in an action for matter coupled to a metric.  Observe that $\cal L\/$
in (\ref{eq:1:15}) is equivalent to a second order Lagrangian for
three fields $\Phi^A = (r^a, \varphi)\/$, with the index $A\/$ being
governed by the metric tensor diag $(1, -1, 1)\/$
\begin{equation}
{\cal L}_{g + m} \to {\cal L}_{\rm equivalent}
           = \frac{\sqrt{-{\frak g}}}{2} {\frak g}^{\mu \nu}
           \partial_\mu \Phi^A \partial_\nu \Phi_A\,.
\label{eq:4:26}
\end{equation}
Here ${\frak g}^{\mu \nu}\/$ is a fictitious metric tensor, unrelated
to the gravitational variables of the theory, but producing the
Lagrange multipliers $u\/$ and $v\/$ in~(\ref{eq:1:15})
\begin{equation}
u = \frac{-\sqrt{-{\frak g}}}{{\frak g}_{11}}\; , \qquad
v = -\frac{{\frak g}_{01}}{{\frak g}_{11}}\,.
\label{eq:4:27}
\end{equation}

Consequently, forming the equivalent action, $I_{\rm equivalent} =
\int d^2 x {\cal L}_{\rm equivalent}\/$, and functionally integrating
$e^{i I_{\rm equivalent}}\/$ over the first two fields, {\it i.e.\/}
over $r^a\/$, produces unity, because of the indefinite metric:
integration of $r^0\/$ involves only the integrand $e^{\frac{i}{2}
\int d^2 x \sqrt{-{\frak g}} \, {\frak g}^{\mu \nu} \partial_\mu r^0
\partial_\nu r^0}\/$ and yields $e^{i I_{\rm effective} ({\frak
g})}\/$, integration of $r^{1}\/$ involves the integrand
$e^{-\frac{i}{2} \int d^2 x \sqrt{-{\frak g}} \, {\frak g}^{\mu \nu}
\partial_\mu r^1 \partial_{\nu} r^1}\/$ and yields $e^{-i I_{\rm
effective} ({\frak g})}\/$.  Thus integration over the two
gravitational variables gives unity and leaves only the matter term
$e^{\frac{i}{2} \int d^2 x \sqrt{-{\frak g}} \, {\frak g}^{\mu \nu}
\partial_\mu \varphi \partial_\nu \varphi}\/$.

\message{frak-mania!}

\item
Observe that our addition $\Delta I\/$ in (\ref{eq:4:17}) is composed
of two terms: one involving $X^{+}\/$, the other $X^{-}\/$.  One may
verify that {\it each\/} term is proportional to the induced Polyakov
action~\cite{ref:17}
$\int \sqrt{-g} R (D^\mu D_\mu)^{-1} R\/$,
in light-cone gauge~\cite{ref:18}.  [Summation formulas
for two Polyakov actions can be given, but they involve a cross
term~\cite{ref:19}, so that the sum occurring in (\ref{eq:4:17}) does
not appear to be expressible in terms of single Polyakov action.]

\item
Finally we remark that the commutator anomaly in (\ref{eq:4:7a}) may
alternatively be cancelled by an addition to $\Theta_\pm$ quite different
from (\ref{eq:4:11}). One easily verifies that
$\Theta_\pm + \alpha_\pm P_\pm' + \beta_\pm X^{\pm''}$ closes with no triple
derivative Schwinger term provided $\alpha_\pm \beta_\pm = \mp (1/48\pi )$.
Owing to the presence of  $P_\pm'$, this modification does not have
the interpretation of a functional connection; in fact it is related
to familiar ``improvements'' of the energy-momentum tensor, which will be
discussed in Section~\ref{sec:newsec}.

\end{enumerate}


\section{Negative energy scalars and states in pure dilaton gravity}
\label{sec:5}

In this Section
we  consider quantization of a scalar field that enters the
Lagrangian with a negative kinetic term [such a field is present in the
dilaton gravity Lagrangian, see (\ref{eq:1:16all})]
and show there are two options. In the first option, standard in
string theory, one works with positive energy states (some of negative
norm) and finds a positive central term, which therefore adds to the one
coming from a conventionally signed Lagrangian, and thus even pure
dilaton gravity, without matter, possesses an obstruction in its
constraint algebra, with resulting center of (+2).

However, we can adopt a second option, where one works with
positive norm states carrying negative
energy.  This leads to a negative central term, which cancels the one
coming from a conventionally signed Lagrangian, thereby removing the
obstruction in the pure dilaton gravity constraint algebra.  This is the choice
made in Section IV.   [Of course, when states are taken to be explicit,
Schr\"odinger representation functionals of the canonical (coordinate)
fields, negative norms cannot be achieved, and only the second option
is available,
as in Section IV.]
In this way we regain the quantization scheme
of Refs.~\cite{ref:9,ref:5,ref:6,ref:7},
which is different from the usual string theory approach. There
are two physical states and we give their explicit form in oscillator
language.
These results can be interpreted as providing a quantization for
a string on two-dimensional Minkowski space,
where {\it all} Virasoro constraints are satisfied,
without introducing ghosts.

\subsection{Indefinite sign quadratic forms and central terms}

Let us consider a free scalar field $\xi(t,\sigma)$
in two space-time dimensions $(t,\sigma)$.
To make rapid contact with string theory results,
the spatial coordinate $\sigma$
is taken on a circle: $\sigma \in [0,2\pi]$,
but this does not alter significantly the infinite-line results
of the previous Sections.
Since we are interested in comparing
scalars with positive and negative kinetic terms,
we let $\xi_+ (\xi_-)$ denote the former (latter).
Henceforth, the $(\pm)$ subscripts do not denote light-cone
components; rather they signal positive $(+)$ or negative $(-)$ kinetic
terms for $\xi_{\pm}$.
In a convenient normalization, the Minkowski action $I^M_\pm$
for both scalar fields  reads
\begin{equation}
I^M_{\pm}= \pm{1\over 8\pi} \int dt \, d\sigma
[ \dot{\xi}_\pm^{\,2} - \xi'^{\,2}_\pm] ~~.
\label{eq:one}
\end{equation}
No additional interactions are included,
because as is seen from equations (\ref{eq:4:6allnew}),
the dynamical problem in our final
version of the pure dilaton gravity model
is governed by quadratic energy and momentum
densities that must annihilate physical states.
The fields in (\ref{eq:one}) correspond
to the gravitational $r^a$ degrees of freedom in (\ref{eq:4:6allnew}).

As usual in path integral quantization, the weight factor is $\exp(iI^M_\pm)$.
The canonical momenta $\Pi_\pm$
and Hamiltonians $H_\pm$ for this system are readily found:
\begin{equation}
\Pi_\pm  = \pm {1\over 4\pi} \dot{\xi}_\pm\,,
\label{eq:twoa}
\end{equation}
\beq
H_\pm = \pm {1\over 2} \int_0^{2\pi} d\sigma
\Bigl( 4\pi \Pi_\pm^{\,2} + {1\over 4\pi} \xi'^{\,2}_\pm \Bigr) ~~.
\label{eq:twob}
\eeq
The basic equal time commutator is
\beq
[ \xi_\pm (\sigma) , \Pi_\pm (\tilde{\sigma})\,]
= i\delta (\sigma-\tilde{\sigma}) \,.
\label{eq:basictime}
\eeq
(A common time argument is suppressed.)
The Hamiltonian equations are solved
by a set of oscillator expansions.
\begin{eqnarray}
\xi_\pm (t,\sigma) &=&
x_\pm + 2t\, p_\pm \, +i \sum_{n\not= 0}\, {1\over n}\,
\Bigl[ \alpha_n^\pm   e^{-in\,(t-\sigma) } +
\overline \alpha^\pm_n e^{-in\,(t+\sigma) }
\Bigr]\,, \nonumber \\
\pm 2\pi\, \Pi_\pm (t,\sigma) &=&  \,\, p_\pm \,\,+ \,\,
\half \sum_{n\not= 0}\, \,
\Bigl[ \alpha_n^\pm e^{-in\,(t-\sigma) }
+ \overline \alpha_n^\pm e^{-in\,(t+\sigma) }
\Bigr] \,,
\label{eq:modemink}
\end{eqnarray}
where hermiticity requires
$({\alpha_n^\pm})^\dagger =\alpha_{-n}^\pm$, and
$({ {\overline\alpha}_n^{~\pm}  })^\dagger =\overline\alpha_{-n}^\pm$.
The non-vanishing commutators
are determined by (\ref{eq:basictime})
\beq
[x_\pm,p_\pm] = \pm i \,,\quad \,
[\alpha^\pm_m , \alpha^\pm_n] = \pm m\,\delta_{m+n,{\scriptscriptstyle 0}}\,\,,
\quad [\ov\alpha^\pm_m , \ov\alpha^\pm_n] = \pm
m\,\delta_{m+n,{\scriptscriptstyle 0}}\,\,.
\label{eq:crel}
\eeq
Note that
the commutation relations for the
oscillators corresponding to the two different scalars differ by a sign.

We define $p_\pm = \alpha_0^\pm = \ov\alpha_0^\pm$,
and take the vacuum state
$\ket{0}$ to be annihilated
by all the operators $\alpha^\pm_n$ and $\ov \alpha^{~\pm}_n$
with $n\geq 0$:
\beq
\alpha^\pm_n \ket{0}=\, \ov \alpha^{~\pm}_n\ket{0}=0\,, \,\, n\geq 0\,.
\label{eq:cftdef}
\eeq
This follows the usual field theory (and string theory) treatment: the
vacuum state is annihilated by the oscillators that appear with
{\it positive} frequencies in the expansion of the field operators,
and this choice is made for {\it both\/} scalars.
An important consequence follows:
the states created by the creation operators have positive energy
for {\it both} scalars. For  scalars with negative kinetic energy,
the norm of some states
will be negative
({\it e.g.} $\alpha^-_{-|n|} | 0 \rangle$), and as will be seen below,
the vacuum is not a localized wave functional in the Schr\"odinger
representation.  Alternatively, another choice of creation and
annihilation operators allows for a localized vacuum wave functional and
gives different conclusions about the structure of the theory;
this will be explained later.

We now continue with the analysis.
It follows from (\ref{eq:modemink}) that at $t=0$
\beq
\half\, (\dot \xi_\pm - \xi_\pm')
= \sum_n \alpha^\pm_n \,e^{in\sigma}\,, \quad
\half\, (\dot \xi_\pm + \xi_\pm')
= \sum_n \ov \alpha^\pm_n \,e^{-in\sigma}\,.
\label{eq:ghty}
\eeq
Furthermore, also at $t=0$
\begin{mathletters}%
\label{eq:getv}%
\begin{eqnarray}%
\pm{\textstyle{1\over 4}}
\, (\dot \xi^2_\pm +  {\xi'}^2_\pm )
&=& \sum_p \Bigl( e^{-ip\sigma} \ov L^\pm_p
+ e^{ip\sigma} L^\pm_p \Bigr) \,,\quad\, \\
\pm{\textstyle{1\over 2}}
\, \dot \xi_\pm \, {\xi'}_\pm \,
&=& \sum_p \Bigl( e^{-ip\sigma} \ov L^\pm_p
- e^{ip\sigma} L^\pm_p \Bigr)\,,
\end{eqnarray}
\end{mathletters}%
where all operators are normal ordered
with respect to the vacuum defined in
(\ref{eq:cftdef}), and
\begin{mathletters}%
\label{eq:defvir}%
\begin{eqnarray}%
L^\pm_p &=& \pm\half\sum_n : \alpha^\pm_{p+n}\alpha^\pm_{-n} : \, , \\
\quad \ov L^{~\pm}_p &=& \pm\half \sum_n : \ov\alpha^\pm_{p+n}
\ov\alpha^\pm_{-n} :\,.
\end{eqnarray}
\end{mathletters}%
These are the Virasoro operators. From
(\ref{eq:twoa}), (\ref{eq:getv}) and (\ref{eq:defvir}), we see
that the Hamiltonians are given by
\beq
H_\pm =  L_0^\pm + \overline L_0^{~\pm} = \pm \left( p_\pm^2
+ \sum_{n=1}^\infty ( \alpha^\pm_{-n}\alpha^\pm_{n} +
\overline\alpha^\pm_{-n}\overline\alpha^\pm_{n}) \right) \,,
\label{eq:hamosc}
\eeq
and the reader can verify that the creation operators indeed increase
the energy both for the case of $\xi_+$ and $\xi_-$.
The Virasoro operators  obey the following commutation
relations
\beq
[ L^\pm_m , L^\pm_n ] = (m-n) L^\pm_{m+n} + {c_\pm \over 12}
(m^3-m)\delta_{m+n,{\scriptscriptstyle 0}}\,\, .
\label{eq:viralg}
\eeq
with the $\ov L^{~\pm}_m$'s satisfying exactly the same commutators. The
central
charge $c_\pm$ is easily obtained from
\beq
{c_\pm \over 2} =
\left\{
{\displaystyle
\bra{0} \, [L^\pm_2 , L^{~\pm}_{-2}] \, \ket{0}
= \bra{0} \, L^\pm_2  L^{~\pm}_{-2} \,\ket{0}
= {\textstyle {1\over 4}} \bra{0} \, \alpha^\pm_1\alpha^\pm_1
\alpha^\pm_{-1}\alpha^\pm_{-1} \,\ket{0}
\atop
\displaystyle
\bra{0} \, [\ov L^\pm_2 , \ov L^{~\pm}_{-2} ] \, \ket{0}
= \bra{0}\, \ov L^\pm_2  \ov L^{~\pm}_{-2} \,\ket{0}
= {\textstyle {1\over 4}} \bra{0} \,
\overline\alpha^\pm_1 \overline\alpha^\pm_1
\overline\alpha^\pm_{-1} \overline\alpha^\pm_{-1} \, \ket{0} }
\right\} ~~.
\label{eq:ccc}
\eeq
We now use the commutation relations (\ref{eq:crel}),
and find that $c_\pm =1$
for the scalars with either positive or negative kinetic term.
While the commutation
relations for the respective oscillators differ by a sign, they must be
used twice, and therefore give the same sign in both cases.

This result can also be presented in the context of the
$[ {\cal E} , {\cal P}]$ commutator.
The energy-momentum tensor for the scalars $\xi_\pm$ is given by
\beq
T_{\mu\nu}^\pm = \pm \half\Bigl( \partial_\mu \xi_\pm \partial_\nu \xi_\pm
- \half \eta_{\mu\nu} (\partial \xi_\pm)^2 \Bigr)\, ,
\label{eq:stens}
\eeq
where $\eta_{\mu\nu} = \hbox{diag}(1,-1)$. It follows that
\begin{eqnarray}
{\cal E}^\pm (\sigma) &\equiv& {1\over 2\pi} T^\pm_{00} (\sigma)
 = \pm {1\over 8\pi}
(\dot \xi^2_\pm +  {\xi'}^2_\pm )
\equiv {1\over 2\pi}\sum_n e^{-in\sigma}{\cal E}^\pm_n \,, \nonumber\\
{\cal P}^\pm(\sigma) &\equiv& -{1\over 2\pi}
T^\pm_{01}(\sigma) = \mp \,{1\over 4\pi}\, \dot \xi_\pm {\xi'}_\pm \,
\equiv - {1\over 2\pi}\sum_n e^{-in\sigma}{\cal P}^\pm_n\,.
\label{eq:stendef}
\end{eqnarray}
Direct comparison with Eq.~(\ref{eq:getv}) gives
\beq
{\cal E}^\pm_n =  \ov L^\pm_n + L^\pm_{-n}  \,,
\quad  {\cal P}^\pm_n = \ov L^\pm_n - L^\pm_{-n}\,.
\label{eq:yes}
\eeq
It is now a simple matter to use (\ref{eq:viralg})
(and its analog for the $\ov L^\pm_n$'s) to find
\beq
[ {\cal E}^\pm_m\, , {\cal P}^\pm_n ] =  (m-n) {\cal E}^\pm_{m+n}
+ {c_\pm \over 6} (m^3-m)\delta_{m+n,{\scriptscriptstyle 0}}\,\, .
\label{eq:viralgg}
\eeq
Since we have shown that $c_\pm =1$, it follows from the above equation
that there is a central term in the
$[ {\cal E}^+_m\,+{\cal E}^-_m\,\, ,\, {\cal P}^+_n +{\cal P}^-_n ]$
commutator. From the definitions (\ref{eq:stendef})
we conclude that, with the quantization scheme described in this
subSection, there is a central term in the
$[ {\cal E}^+(\sigma) +{\cal E}^-(\sigma)\,\, ,\,
{\cal P}^+(\sigma) +{\cal P}^-(\sigma) ]$ commutator of the total energy
and total momentum densities. Thus the combined system of a positive norm and
a negative norm scalar exhibits a central term in the relevant
commutator, so that the total ${\cal E}$ and ${\cal P}$
operators cannot annihilate a state.

\subsection{Another choice}

Since the Lagrangian with negative kinetic term gives rise to a
Hamiltonian that appears negative (in the above treatment positive
energies are nevertheless achieved at the expense of negative-norm
states), one could say that the annihilation operator corresponds to a
{\it negative} frequency oscillator in the expansion of field operators.
Thus, in contrast to the usual conformal field theory choice
(\ref{eq:cftdef}), we can take
\begin{eqnarray}
{\alpha^+_n}\ket{0}&=&\, {\ov \alpha^{~+}_n}
\ket{0}=0\,, \,\, n\geq 0 ~~,
\nonumber\\
{\alpha^-_n}\ket{0}&=&\, {\ov \alpha^{~-}_n}
\ket{0}=0\,, \,\, n\leq 0 ~~,
\label{eq:cftdeff}
\end{eqnarray}
and now there are no negative norm states.

A further argument in favor of (\ref{eq:cftdeff}) can be made.
Note that the positive frequency modes are projected by
\beq
\int_0^{2\pi} d\sigma \, e^{-i\sigma n }
\left(
\pm \Pi_\pm (t,\sigma) - {i |n| \over 4\pi} \, \xi_\pm (t,\sigma)
\right)
=
\cases {\alpha_n^\pm \, e^{-int} & $n>0$ \cr
        \bar{\alpha}_{|n|}^\pm \, e^{-i|n|t} & $n<0$\cr}
\label{eq:onetwenty}
\eeq
This may also be presented with the aid of the positive kernel
\beq
\omega(\sigma, \tilde \sigma ) = {1\over 8\pi^2} \sum\limits_{n}
e^{in(\sigma-\tilde\sigma) } |n| \,,
\label{eq:onetwentyone}
\eeq
\beq
\int_0^{2\pi} d\sigma \, e^{-i\sigma n}
\left( \pm \Pi_\pm (0, \sigma) - i
\int_0^{2\pi} d \tilde{\sigma} \, \omega(\sigma , \tilde{\sigma})
\, \xi_\pm (0,\tilde{\sigma}) \right) =
\cases {\alpha_n^\pm & $n>0$\,, \cr
        \bar{\alpha}_{|n|}^\pm & $n<0$\,. \cr}
\label{eq:onetwentytwo}
\eeq
When these modes are taken to annihilate the vacuum,
as in conformal theory [see (\ref{eq:cftdef}) ],
we have in the Schr\"odinger representation
\beq
\left\{
\pm {\delta \over \delta \xi_\pm (\sigma)}
+ \int_0^{2\pi} d\tilde{\sigma} \,
\omega(\sigma , \tilde{\sigma})
\, \xi_\pm (\tilde{\sigma}) \right\}
\Psi_\pm^0 (\xi_\pm) = 0\, ,
\label{eq:onetwentythree}
\eeq
where $\Psi_\pm^0$ is the vacuum functional depending on
$\xi_\pm(\sigma) \equiv \xi_\pm (0,\sigma)$.
The unique (up to normalization) solution is
\beq
\Psi_\pm^0 (\xi_\pm) = \exp \Bigl(
{\mp {1\over2} \int \xi_\pm \,\omega\, \xi_\pm}\Bigr) ~~.
\label{eq:onetwentyfive}
\eeq
We see that for $\xi_-$, whose kinetic term is negative,
the functional grows inadmissibly as a quadratic exponential, since
the kernel is positive.   On the contrary if in this case the modes
annihilating the vacuum are taken with negative frequency, one obtains
an acceptable Gaussian, as in the usual case with positive kinetic term
[see (\ref{eq:4:7new})].

With the choice (\ref{eq:cftdeff})
one readily verifies that states created with the
positively moded oscillators have positive norms, but
negative energies.
It also follows that the central term $c_-$ corresponding to
the  scalar with negative kinetic term, quantized in this way,
is given by $c_-=-1$.
Indeed, the computation in (\ref{eq:ccc})
changes by a sign since this time $L_{-2}$
annihilates the vacuum \cite{ref:flor}.

\medskip
With the ordering prescription we are considering now,
there is no central term in $[{\cal E}^+(\sigma) + {\cal E}^-(\sigma),
{\cal P}^+(\tilde{\sigma}) + {\cal P}^-(\tilde{\sigma})]$,
so that the total ${\cal E}$ and ${\cal P}$
operators can annihilate a state. The relevant states, in the
Schr\"odinger representation, read $\exp ({\pm {i\over 4\pi}
 \int d\sigma \, \xi_+ \xi'_-})$
[compare (\ref{eq:4:5new})], and we seek their Fock space equivalents
to verify explicitly
that they are annihilated by all the Virasoro conditions. There is
a well-defined procedure to go from a wave functional to its corresponding
state, and coherent state methods are most efficient.
Since these methods are familiar, we just give the result, which follows
after a certain amount of simple computation.
The relation of the above state to the Fock vacuum is given by
[compare (\ref{eq:4:8new})]
\beq
| \pm \rangle = \Bigl\{ \prod_{n=1}^\infty \, 2\cdot
\exp \Bigl( \pm  {1\over n}\Bigl[
\alpha_{-n}^+\alpha_n^- -
\overline\alpha_{-n}^{~+}\overline\alpha_n^{~-}
\Bigr]\Bigr) \, \Bigr\} \,| 0 \rangle\,\,.
\label{eq:onetwentysix}
\eeq
Recall that for the positive signature scalar the negatively moded
oscillators are creation operators and for the negative signature
scalar the positively moded oscillators are creation operators. It
follows that all the oscillators appearing in the above exponential
are creation operators, as they should be for a normal ordered
representation of the state. It is interesting to note that
the straightforward transcription from the Schr\"odinger representation
gives a state with the factor of two shown above, which prevents the
state from being a linear superposition with finite coefficients
of Fock space states (states built
with a finite number of oscillators acting on the vacuum).
We can drop this factor
and concentrate on the nontrivial part of the states
\beq
| \Psi_\pm \rangle =
\exp \Bigl( \pm \sum_{n=1}^\infty {1\over n}\Bigl[
\alpha_{-n}^+\alpha_n^- -
\overline\alpha_{-n}^{~+}\overline\alpha_n^{~-}
\Bigr]\Bigr)  \,| 0 \rangle\,\,.
\label{eq:thirdadd}
\eeq

The Virasoro conditions or physical state conditions demand that
the operators ${\cal E}^\pm_n$ and ${\cal P}^\pm_n$ given in
(\ref{eq:yes}) must annihilate $| \Psi_\pm \rangle$ for all values of $n$.
By taking suitable linear combinations of these constraints they can
be put in the form  $ {\cal L}_m | \Psi_\pm \rangle
= \overline{\cal L}_m | \Psi_\pm \rangle = 0$ for all $m$, where the operators
\beq
{\cal L}_m \equiv  L_m^{+} + L_m^{-} \,, \quad
\overline {\cal L}_m \equiv \overline L_m^{~+} +\overline L_m^{~-} \,,
\label{eq:fourthadd}
\eeq
are the total Virasoro operators of the system, and the relevant
oscillator expansions are given in
(\ref{eq:defvir}). Because holomorphic and antiholomorphic sectors
of the constraints decouple, it is sufficient to verify that
\beq
{\cal L}_m \,e^{\pm \Omega} \, | 0 \rangle = 0, \quad
\hbox{where}\quad
\Omega = \sum_{n=1}^\infty {1\over n} \alpha_{-n}^{+} \alpha_n^{-} \,.
\label{eq:fifthadd}
\eeq
Since the Virasoro operators have at most two annihilators and $\Omega$
has two creators, multiple commutators with more than two $\Omega$'s
must vanish. It is therefore enough to verify that
\beq
\Bigl( {\cal L}_m + \, [{\cal L}_m , \pm \Omega] +
{1\over 2} [ \, [ {\cal L}_m , \Omega]\, ,\Omega ] \Bigr) |0\rangle =0\,.
\label{eq:sixthadd}
\eeq
One readily checks that the above equation is satisfied since
$ [{\cal L}_m , \pm \Omega] |0\rangle =0$,
and the other two terms cancel each other. As a consequence
(\ref{eq:fifthadd}) holds, and  the states
$|\Psi_\pm \rangle$ satisfy {\it all} the Virasoro constraints.

This result can be interpreted as providing physical states in a
quantization of a string on two-dimensional Minkowski space without
introducing ghosts.   To be sure, the state space for
this string, being two-fold, is rather small.  Nevertheless, it is
interesting that {\it all\/} Virasoro constraints can be satisfied, and
of course in the gravitational context, the states give an adequate
description of pure dilaton gravity on a line \cite{ref:9,ref:5,ref:6,ref:7}.
A similar construction can also be carried out for a string in
$(d,d)$ dimensional space-time.

A final observation is in order. Since the constraints decouple the holomorphic
and antiholomorphic sector, we actually find four states satisfying the
constraints: the extra two corresponding to the remaining sign combinations
that can be constructed in (\ref{eq:thirdadd}). Nevertheless,
one can show that such states do not have associated wave functionals;
the procedure for obtaining wave functionals from states does not work
due to infinities. We therefore suspect that these other two states are
not likely to be relevant. There may be other physical states, but we have
not found them. (The strategy of multiplying $|\Psi_\pm\rangle$
by an operator that
commutes with all Virasoro operators does not give anything new.)

\section{Achieving zero center within the Conformal Approach}
\label{sec:newsec}

In this Section we go back to the  quantization choice for the
negative signed scalar field conventional in conformal theory, so that
its center is $c_-=1$. We use
the Euclidean treatment with $z\equiv\tau+ i\sigma, \, \bar z
\equiv \tau -i\sigma$, ($\partial\equiv {\partial\over \partial z},$
$\overline\partial\equiv {\partial\over \partial\bar z}$)
but retain indefinite signature for the two gravitational scalar
fields.  We discuss
``improvements'' of the energy-momentum tensor and then show how
to achieve constraints without a center.

\subsection{Central terms and improvements}

Let $\xi_+$ and $\xi_-$  denote  free
scalar Euclidean fields with  positive and negative kinetic terms respectively.
For both scalars we set $\xi_\pm (z,\bar z) = \xi_\pm (z)
+ \bar \xi_\pm (\bar z)$, as implied by the equation of motion, and one has
\beq
\langle \xi_\pm (z_1) \xi_\pm (z_2)\rangle = \mp \ln (z_1-z_2)\,, \quad
\langle \bar \xi_\pm (\bar z_1) \bar \xi_\pm (\bar z_2 )\rangle =
\mp \ln (\bar z_1-\bar z_2)\, . \label{eq:noname}
\eeq

We shall now discuss the standard improvement terms that can change the
central charge of a two-dimensional scalar field. Consider the
holomorphic component of the energy-momentum tensor,
improved by a linear term
\beq
T_\pm  (z) =  \mp \half \partial \xi_\pm (z) \partial \xi_\pm (z)
 + {Q_\pm\over 2}
\partial^2 \xi_\pm (z)\, ,\label{eq:impst}
\eeq
where we take the constants $Q_\pm$ to be real (in Minkowski
space this makes the extra term in the energy-momentum tensor real).
A short calculation gives
\beq
 c_{\pm} = 1 \pm 3Q^2_\pm \,.\label{eq:ctimpt}
\eeq
Therefore the improvement increases the central charge of
the positively signed scalar $\xi_+$,
and decreases the one of the negatively signed scalar $\xi_-$.
The constants $Q_+$ and $Q_-$ are arbitrary. Choosing them allows setting
the total central charge to any desired value, and various choices will
be made below. For definiteness we shall take the choice where $Q_+$ or
$Q_-$ vanishes, and the other is adjusted to fix the total central charge.

For later reference we note that when
the (holomorphic) energy-momentum tensor
for a free Bose field $\chi (z, \bar z)
= \chi (z) + \overline \chi (\bar z)$
takes the form
\beq
T (z) =  - \half \partial \chi (z) \partial \chi (z) + {Q\over 2}
\partial^2 \chi (z)\, ,\label{eq:impstt}
\eeq
then the local operators $e^{w\chi(z)}$ have dimension
\beq
\hbox{dim}\,\Bigl( e^{w\chi(z)} \Bigr) = {Q^2\over 8} - {1\over 2}
\Bigl( w - {Q\over 2} \Bigr)^2\, .\label{eq:dim}
\eeq
Moreover, since the translation current $i\partial\chi$ is no longer
a tensor when $Q\not= 0$, the state-operator correspondence  for
momentum eigenstates is not the standard one \cite{ref:polchinski}.
\beq
\ket{p} \,\Longleftrightarrow \, \exp \Bigl( ip\chi (z) + {Q\over 2}\chi (z)
\Bigr) \,.\label{eq:shift}
\eeq
It follows from the last two equations that
\beq
L_0 \ket{p} = \Bigl( {Q^2\over 8} + {p^2\over 2} \Bigr)\ket{p}\, .
\label{eq:uff}
\eeq

\subsection{Achieving zero center}

When we quantize pure dilaton gravity (without matter but with
cosmological term) in the way conventional for conformal
theory, we obtain a system with central charge
$c=2$, and there is an obstruction.
The (holomorphic) energy-momentum tensor reads
\beq
T  = -\half \partial \xi_+\partial \xi_+ + \half \partial \xi_-\partial \xi_-
\,\, .\label{eq:stensr}
\eeq
We improve the negative norm scalar to $c_-=-1$ by taking
$Q_-= \sqrt{2/3}$. Thus
\beq
\widetilde T  = -\half \partial \xi_+\partial \xi_+
+ \half \partial \xi_-\partial \xi_-
+{1\over \sqrt{6}} \partial^2 \xi_-\,, \label{eq:newimpx}
\eeq
defines constraints with no obstruction. There is no guarantee, however,
that there are interesting states satisfying the constraints. We suspect
that the physical states, if any, are not  finite linear superpositions
of Fock space states, but rather infinite linear superpositions of such
states [as is the case for pure dilaton gravity, quantized so that
obstructions cancel, see (\ref{eq:thirdadd})].

Let us now include one free
matter field, $\varphi (z,\bar z) = \varphi (z) + \overline\varphi (\bar z)$.
The relevant energy-momentum tensor is then
\beq
T  = -\half \partial \xi_+\partial \xi_+\
\,\,+\, \half \partial \xi_-\partial \xi_- -\half \partial \varphi
\partial \varphi
\,. \label{eq:newipx1}
\eeq
This produces a central term $(+3)$
and the resulting constraints cannot
be consistently imposed.
Again, we improve in the negative norm scalar sector to
lower the central charge of that scalar, and get zero for the total center.
Taking $Q_-=1$ we obtain
\beq\widetilde T  = -\half \partial \xi_+\partial \xi_+\
\,\,+\,  \half \partial \xi_-\partial \xi_- + \half \partial^2 \xi_-
 -\half \partial \varphi \partial \varphi
\,, \label{eq:newime}
\eeq
This energy-momentum tensor presents no obstruction to
quantization, but again, it
remains to be seen whether there are interesting physical states.

Finally we note that in the presence of matter, with gravity quantized
as in Section~\ref{sec:4}, where the negative signed
gravitational scalar cancels the
center of the positive signed gravitational scalar, there remains the center
$c=1$ from the matter field. This may be removed by improving
the energy-momentum tensor of the negatively signed scalar as is indicated in
point (5) at the end of Section~\ref{sec:4}. In the present formalism this
corresponds to $Q_-= \sqrt{2/3}$.

\section{BRST quantization of the dilaton gravity theory}
\label{sec:6}

In this Section we consider the quantization of
dilaton gravity by the covariant (BRST) method. Conformal
field theory conventions are followed for the negative-signed
scalar, so that $c_-=1$. To begin one must add
the reparameterization ghost and
antighost, as is justified by noting that
the original action is invariant under two-dimensional
diffeomorphisms, and BRST quantization  requires that the
two-dimensional Lagrangian be supplemented by gauge fixing and
ghost/antighost terms.  Upon selecting the conformal gauge,
the Weyl factor of the
two-dimensional metric (Liouville field) together with the dilaton field
comprise the two gravitational variables.

In two-dimensional gravity-matter theories,
consistent quantization follows when
the combined gravity,
matter and ghost degrees of freedom define a conformal
field theory with vanishing total central charge \cite{ref:distlerkawai}.
Since the ghost conformal field theory gives a contribution of $(-26)$
to the central charge, the
gravity and matter degrees of freedom must
give a total central charge of $(+26)$ and this will require changing
the central charge by adding a background charge to the energy-momentum
tensor of the positively-signed scalar.

This Section has three parts. In the first and second parts, we set
up the formalism for pure dilaton gravity and for matter-coupled
dilaton gravity
respectively. In the third part, we discuss physical states and the
semi-classical limit.

\subsection{Pure Dilaton Gravity}

In the BRST approach to quantization, we improve the
positive signed scalar $\xi_+$ to reach total central charge $(+26)$ for
both scalars. Then the
theory is coupled to the ghost system which has total central charge $(-26)$.
We therefore need
$Q_+= 2\sqrt{2}$ leading to $c_+=25$, $~c_++c_- = 26$.
The relevant (holomorphic)
energy-momentum tensor is
therefore given as
\beq
\widetilde T (z) =  T_{\hbox{gr}}(z) +
T_{\hbox{gh}}(z),\label{eq:neweqn}
\eeq
where the improved gravity  tensor $ T_{\hbox{gr}} (z)$
and the ghost tensor
$ T_{\hbox{gh}} (z)$  are
\begin{eqnarray}
 T_{\hbox{gr}}
 &=& -\half \partial \xi_+\partial \xi_+ +\sqrt{2} \partial^2 \xi_+
\,\,+\, \half \partial \xi_-\partial \xi_-\, ,\label{eq:yyp}\\
 T_{\hbox{gh}}  &=& 2(\partial c) b + c
\partial b \,. \label{eq:yyy}
\end{eqnarray}
Here $c$ and $b$ are the holomorphic ghost and antighost fields respectively,
and satisfy $\langle b(z) c(w)\rangle = 1/(z-w)$.
The total energy-momentum tensor $\widetilde T$ has no central term.
In this quantization procedure
physical states
are annihilated by
the BRST operator
\begin{mathletters}%
\begin{eqnarray}
Q_B &=& \oint dz \, c(z) \, (T_{\hbox{gr}}(z) +
\half  T_{\hbox{gh}}(z)) + {\scriptstyle\rm antiholomorphic} ~~, \nonumber\\
Q_B \, \ket{\hbox{phys}} &=& 0 ~~.
\end{eqnarray}
Physical states satisfy an equivalence relation (cohomology),
\begin{equation}
  \ket{\hbox{phys}} \equiv \ket{\hbox{phys}} + Q_B \ket{\alpha} ~~.
\end{equation}
\end{mathletters}%
Since, on general grounds $\{ Q_B , b\} = \widetilde T$,
it follows that physical states do not
have to be annihilated by $\widetilde T$.
One just has
\beq
\widetilde T \, \ket{\hbox{phys}} = \{ Q_B , b\} \,\ket{\hbox{phys}}
=   Q_B \Bigl(  b\,\ket{\hbox{phys}}\Bigr) \,,
\eeq
namely, $\widetilde T$ on physical states must give a BRST trivial state
(almost zero!). It follows that the states  annihilated by
$\widetilde T$ do not coincide with the BRST physical
states. While the states annihilated by $\widetilde T$ are likely to be
very few, the spectrum of BRST physical states is quite rich.
In some cases, the BRST physical states are in correspondence with
the states annihilated by the {\it positive} modes  of the matter
part of the energy momentum tensor, the so-called ``matter primaries''.

The total
energy-momentum tensor in (\ref{eq:neweqn})-(\ref{eq:yyy})
coincides  with that of the
``Gaussian model'', a model where a single boson is minimally
coupled to pure two-dimensional gravity (frequently called
``$c=1$'' model, highlighting the matter contribution to the central term.).
The field $\xi_+$ plays
the role of the Weyl factor in the metric
tensor (Liouville field), and $\xi_-$ plays the role of the single
boson. In the Gaussian model, however, this boson has positive kinetic
energy, so its contribution to $T_{\hbox{gr}}$
enters with opposite sign in (\ref{eq:yyp}).

\subsection{Matter-coupled dilaton gravity}

Let us now include a single
matter field,  $\varphi$. The relevant energy-momentum tensor is then
\beq
T  = -\half \partial \xi_+\partial \xi_+\
\,\,+\, \half \partial \xi_-\partial \xi_-
-\half \partial \varphi \partial \varphi
\,. \label{eq:newipx2}
\eeq
In the BRST approach, we improve
$\xi_+$ to get $c_+=24$ and introduce the ghosts.
This requires $Q_+ =\sqrt{23/3}$
\beq
\widehat T  = -\half \partial \xi_+\partial \xi_+
\,+\,\sqrt{{23\over 12}}\, \partial^2 \xi_+\,\,  + \half
\partial \xi_-\partial \xi_-  -\half\partial\varphi\partial \varphi
 \, + 2(\partial c) b + c\partial b\,.
\label{eq:newimp}
\eeq
Since this energy-momentum tensor has zero central term one can now
find physical states within
BRST quantization. We shall discuss the physical states
of this theory in the next subSection.

Yet another approach would
begin by interpreting the starting point for quantization
as a system having three scalar fields  $(\xi_+,\xi_-, \varphi)$
on a curved world-sheet. Quantization would proceed
by adding a further gravitational  field $\varphi_g$ with $c_g = 23$, and
the $(b,c)$ ghost system. Physical states would be defined by the
corresponding BRST operator. We shall not discuss this possibility
any further.

\subsection{Physical states and a comparison with the flat space spectrum}

In this subSection we discuss physical states in the BRST
quantization of pure dilaton gravity,
and matter-coupled dilaton gravity.
To understand the pure dilaton gravity case we begin by discussing the
Gaussian model, whose physical states \cite{ref:goldstone}
are in correspondence with those of pure dilaton gravity.
We then examine a question relevant to the semi-classical
limit of the BRST quantized matter-coupled dilaton gravity theory.
We use the results of Ref.~\cite{ref:bilal} to enumerate the
physical states, and show that the spectrum differs significantly
from the spectrum of a free massless particle propagating
in flat space in the absence of gravity.

\medskip
When pure two-dimensional gravity (no dilaton)
is coupled minimally to a free massless positive signature boson $X$,
one obtains the Gaussian model. This model can
be viewed in two different ways: as a quantum gravity
field theory in two dimensions
with very few physical states, or as a string theory whose target space
is two dimensional. Both theories arise as different interpretations for the
BRST quantization of the Gaussian model.

  The BRST quantization  of the Gaussian model proceeds by
adding the $(b,c)$ ghost system to account for diffeomorphism invariance,
and using a positive signature Liouville field
$\varphi_L$
to represent the Weyl factor of the metric.
A background charge $Q_L= 2\sqrt{2}$ is included for $\varphi_L$, and thus
its center is $c_L= 25$.
Together with the field $X$, $\varphi_L$ forms a system
of center $c=26$, the right amount to be cancelled
by the ghost system.
The physical states of such system, defined
by the  BRST cohomology at the relevant ghost number, are well known
\cite{ref:goldstone}. Among the physical
states, there is one family parametrized by a continuous parameter. It arises
from consideration of the  state
\beq
|V(p_X, p_L) \rangle = c_1\bar{c}_1 \,| p_X, p_L\rangle
\,\,.\label{eq:ctachdef}
\eeq
where $c_1$ and $\bar c_1$ are oscillator modes of the fields $c(z)$ and
$\bar c(\bar z)$, and $| p_X, p_L\rangle$ is a vacuum state carrying momenta
$p_X$ and $p_L$ in the matter and Liouville sectors respectively.
For such a state, the BRST condition simply requires that
$L_0|V(p_X, p_L) \rangle = \overline L_0 |V(p_X, p_L) \rangle =0$.
Explicitly these conditions give
\beq
L_0|V(p_X, p_L) \rangle = \Bigl(-1 + {p_X^2\over 2}
+ 1 + {p_L^2\over 2}    \Bigr) c_1\bar{c}_1 \,| p_X, p_L\rangle = 0\,,
\label{eq:pctach}
\eeq
and exactly the same equation for $\overline L_0$.
In the above equation we have indicated explicitly the contributions to
$L_0$ coming from the ghost oscillators $(-1)$, from the matter
(${1\over 2} p_X^2$), and from the Liouville sector
($1 + {1\over 2} p_L^2$),
as follows from Eq.(\ref{eq:uff}).
Therefore, the physical state conditions require
\beq
p_X^2 + p_L^2 = 0 \, \rightarrow\quad  p_X = p \,, \quad p_L = \pm ip \,,
\label{eq:solcon}
\eeq
where $p$ is an unconstrained real parameter, and the physical states are
given as
\beq
|V_\pm (p) \rangle = c_1\bar{c}_1 \,| p, \pm ip\rangle\,\,.
\label{eq:tstate}
\eeq
In addition to this state, parametrized by a continuous variable $p$, one
finds states of the type
\beq
(\alpha_{-m_1}\cdots\alpha_{-m_p}\varphi_{-k_1}\cdots\varphi_{-k_q})
(\bar\alpha_{-n_1}\cdots \bar\alpha_{-n_s}\bar\varphi_{-l_1}
\cdots\bar\varphi_{-l_t})
c_1 \bar c_1\ket{p_X, p_L}\,\,, \label{eq:discrstate}
\eeq
obtained by acting with some number of creation operators of $X$ ($\alpha_{-n}$
and $\bar \alpha_{-n}$) and $\varphi_L$ ($\varphi_{-n}$ and
$\bar \varphi_{-n}$)
on the momentum eigenstate $\ket{p_X, p_L}$.  Here the BRST conditions
force $p_X$ and $ip_L$ to be fixed real numbers. The BRST conditions also fix
specific combinations of creation operators.
Due to the absence of a continuous parameter in the allowable momenta
these are called {\it discrete states}. (There
are also more complicated discrete states involving ghost oscillators.)

We can use the BRST quantization of the Gaussian model as a quantization
of pure dilaton gravity. The cohomology problem, which determines physical
states, is the same for the two models once we identify the two dilaton
gravity fields $\xi_+$ and  $\xi_-$ with $\varphi_L$ and $X$ respectively,
except for the change of signature needed to replace $\xi_-$ with $X$.
This change is of little relevance. It is possible to use the results
of Bouwknegt.{\it et.al.}\cite{ref:goldstone}
to see that the set of BRST physical
states is not altered except for the changes implied by the
signature-altering identification  $X \leftrightarrow i\xi_-$.
Notice that in this way we obtain a
set of physical states for pure dilaton gravity
which is far richer than the two states gotten by non-BRST
quantization methods described in Section\ref{sec:5}. At the same time the
number of states is not large enough to give rise to a propagating
field degree of freedom (see below). This is consistent with expectations from
a naive degrees of freedom count: pure two-dimensional gravity giving
$3-2\times 2=-1$, and the dilaton counting as $+1$, for a total of zero.

[As a digression let us remark that it makes more sense to interpret the
BRST-quantized Gaussian model as pure dilaton gravity, rather than as
pure gravity (no dilaton) coupled to a massless {\it matter} field.
The point is that after BRST quantization almost nothing is left of the
scalar field ~-- there are too few quantum states. Indeed the set of
BRST physical states bears no resemblance to the set of
physical states that would be obtained by quantizing a scalar massless
field on a  flat two-dimensional space without gravity.
If the space coordinate is a circle, the physical states
of a single two-dimensional free massless boson can be obtained
from the expansion
of $\xi_+$ given in (\ref{eq:modemink}), and are of the form
\beq
(\alpha_{-m_1})^{n_1}\cdots (\alpha_{-m_p})^{n_p}
(\bar\alpha_{-l_1})^{r_1}\cdots (\bar\alpha_{-l_q})^{r_q}\,
\ket{\, p}\,,\label{eq:fbextp}
\eeq
where $p\in [-\infty , +\infty]$ is an arbitrary constant labeling the
eigenvalue of the $\Pi_+$ zero mode. These
states, arising from quantization on a cylinder, are the analogs of
the many-particle states obtained by quantizing on the
infinite space-like line.
The state with no oscillators  may be paired naturally with
the state given in (\ref{eq:tstate}), but the states with oscillators
in (\ref{eq:fbextp}) still have a continuous
parameter. That parameter is absent in the discrete BRST states
(\ref{eq:discrstate}), moreover, the discrete states are anyway too few.
(All this is consistent with the situation in the unquantized, classical
theory where the equation for the matter field $X$
that follows from varying
$g_{\mu\nu}$ requires the vanishing of the matter energy-momentum tensor
$\theta_{\mu\nu} = \partial_\mu X\partial_\nu X -\half g_{\mu\nu}
g^{\alpha\beta} \partial_\alpha X \partial_\beta X$, which in turn means
either that $X$ is constant and $g_{\mu\nu}$ undetermined,
or $g_{\mu\nu}$ is singular.)]

The BRST-quantized Gaussian model may alternatively
be used to define a two-dimensional string theory.
In this case $\varphi_L$ and $X$ are viewed as coordinates for
a two-dimensional target space, the space-time of the string theory.
Following the standard paradigm, to
each state of the Gaussian model one assigns a space-time field, and
reinterprets the physical state conditions as linearized equations of
motion together with gauge conditions for the  space-time field.
In doing this one must decide whether $\varphi_L$ or $X$ is to
be analytically continued to represent time. Both choices have been made in
the literature. We shall choose $X$, and thus
we let $p_X \to ip_X$. In our present context this is suggested from
the correspondence with dilaton gravity where the negative signature
field $\xi_-$ is identified with $X$. Following the string paradigm,
associated
to the state (\ref{eq:tstate}) we  have a
scalar field $\eta  (p_X , p_L)$, whose linearized
equation of motion requires $-p_X^2 +p_L^2=0$. This is therefore a massless
scalar field, somewhat misleadingly called the ``tachyon''. This
is the only  quantum field in the theory. The discrete states
lead to space-time ``fields'' whose linearized equations of motion and gauge
conditions are so strong that they eliminate all degrees of freedom except for
a single constant: the value of the field at the allowed momenta. These
degrees of freedom are parameters of the background
space-time. Sigma model studies have shown that the effective field theory
limit of this string theory is precisely the two-dimensional dilaton gravity
theory indicated in Eq.(\ref{eq:1:2}) coupled to a single massless field.
To summarize the indirect argument:  pure dilaton gravity theory,
after transformation to quadratic form, is BRST quantized as a
Gaussian model; then a string theory is constructed, which in the
effective field theoretic limit reproduces a dilaton gravity model
coupled to a single massless field.

We can now proceed with the case of dilaton gravity coupled minimally
to a single matter field $\varphi$.
The energy-momentum tensor relevant for the BRST quantization of this theory
was given in Eq.(\ref{eq:newimp}). In order to compare this system
to those in the literature, it is convenient to
label $(\xi_- ,\varphi ) \equiv (X^0, X^1)$ as `matter' fields,
and to label $\xi_+\equiv \varphi_L$ as the Liouville field .
The energy-momentum tensor
then reads (with $Q= \sqrt{23/3}$)
\beq
\widetilde T (z) = -\half \partial \varphi_L\partial \varphi_L\
\,+\,\sqrt{23\over 12}\, \partial^2 \varphi_L \,\,+ \half
\partial X^0\partial X^0  -\half \partial X^1 \partial X^1
\, \, + 2(\partial c) b + c\partial b\,.
\label{eq:newt}
\eeq
The general results of Ref.\cite{ref:bilal}
concerning BRST cohomology for $d$ matter fields coupled to a Liouville
field apply for the case when $d=2$. For the standard ghost number, and
apart from some discrete states, the result
is simple. The cohomology is isomorphic to a
$(d-1)$ dimensional on-shell Fock space, and is therefore one-dimensional
for our present case. While the results of
Ref.\cite{ref:bilal}
were given for the holomorphic sector, the total cohomology for the
two sectors combined is just the tensor product, except for discrete states.
Thus the Fock space must include both barred and unbarred oscillators.
Letting $( p^0,p^1,p_L)$ denote the zero modes of $( X^0, X^1, \varphi_L )$,
we can describe the counting of states
in the cohomology as
the set of states obtained by acting with a complete
single set of oscillators $(a^\dagger_{n}, \overline a^\dagger_{n})$ ($n>0$) on
a vacuum $\ket{p^0,p^1,p_L}$, where the on-shell condition arising from
the $(L_0+\overline L_0)=0$ constraint requires that
\beq
 -(p^0)^2 + (p^1)^2 + (p_L)^2
+{Q^2\over 4} + N -2 =0\, .\label{eq:onshell}
\eeq
Here $N\geq 0$ is the total number operator of the oscillators that
are acting on the vacuum state.

At first sight the field theoretical interpretation of the BRST cohomology
seems encouraging since we saw that the free scalar field without
gravity has precisely a Fock space that is generated with a single complete
set of oscillators. The zero modes, however, do not seem to work out.
In the case of one matter field with no gravity, the zero mode $p$
labeling the vacuum amounts to one single parameter.
Here, in dilaton gravity, we have three zero modes
$(p^0,p^1,p_L)$ and only one constraint, thus two parameters.
This mismatch is generic, and will not  disappear by adding  extra
matter fields. We therefore conclude that there is a difference
between the naive semi-classical spectrum we would expect for a free scalar
field and the spectrum we have obtained by BRST quantization of
matter coupled dilaton gravity. (A string interpretation of the
BRST physical states would correspond to a three dimensional target
space and is not relevant to our purposes.)

If this difference is truly significant, and remains so when the space
is not a circle but an open line, one would reach the conclusion that
there seems to be no known two-dimensional
quantum gravity theory coupled to matter whose set of physical states
resembles closely that of a matter theory without gravity.
A possible exception to this conclusion might be the quantum gravity
provided by two-dimensional string theory. In this case there is strong
evidence that the semi-classical limit of this theory corresponds to
a massless scalar field propagating on a particular two-dimensional
space-time background \cite{ref:polchinski}.

\section{Conclusion and Discussion}
\label{sec:8}

In this paper we have addressed questions relevant to the
definition and quantization of two-dimensional dilaton gravity.
One important issue concerns
field redefinitions and the starting point for quantization.
In Ref.\cite{ref:bilalcallan} a series of field redefinitions, some
not invertible,
reduced the dilaton gravity theory {\it without} cosmological constant to a
conformal theory of two free scalar fields
with opposite norm. As we have seen, even the theory {\it with}
non-zero cosmological term can be brought to the same final form, as was
done in Ref.\cite{ref:9} and  Section~\ref{sec:3} with the help of
temporally local but spatially non-local field transformations.
An obvious question is whether the physics is changed by such
redefinitions, and the fact that the same final theory emerges,
regardless whether a cosmological constant is present, suggests
that field redefinitions are non-trivial in the quantum case.
Perhaps the redefinitions ought to be interpreted as part of the
definition of the quantum theory;
namely, the classical system is first manipulated into a particular form,
and then quantized.

After carrying out our reduction to indefinite quadratic form
we have seen how pure dilaton gravity can be quantized without
obstruction by adopting  particular quantization scheme for the
scalar with negative kinetic energy.
As a byproduct, we found a unitary quantization of a string in
two-dimensional Minkowski space where {\it all\/}
Virasoro constraints are satisfied. There are only two physical states,
and therefore this is not a rich spectrum.
Nevertheless, as a matter of principle, it is quite surprising that there
are nontrivial states annihilated
by all the Virasoro operators, since in standard string theory
the vacuum state and all physical states are only annihilated by the positively
moded Virasoro operators.  While our states have infinite norm
they can be presented as well defined linear superposition of Fock space
states.

Once matter is included, the theory is still of indefinite quadratic form, but
the constraints cannot be solved owing to the quantal commutator anomaly.
We emphasize that classically the constraints
can be satisfied and the equations
can be solved \cite{ref:9}. Also when the quantized matter comprises
point particles, there are no serious obstructions to satisfying the
constraints and the quantum theory can be analyzed to the end
\cite{ref:bakseminara}.
It is only quantized matter fields that produce problems. We have seen that
progress can be made when various modifications or improvements are effected.
While the obstructions can be overcome in various ways,
the physical picture in the resulting
quantum theory bears no resemblance to the classical physics, except in
BRST quantization where there is a vague resemblance. This suggests
that semi-classical
analysis may be of questionable relevance  even for an approximate
description \cite{ref:guilini,ref:mathur}.

It would be  interesting to clarify the status of  quantization
of four-dimensional gravity in  light of the observations
made in the present work.  Since physical gravity has propagating
states, we would expect that the relevant constraints of the quantum
theory will exhibit a center even in the absence of matter.
If this center could be determined
precisely one could investigate if it can be removed
by introducing special variables, improvement terms, and/or by
the use of alternative quantization schemes, as in the two-dimensional
model.

\medskip
\noindent
\underbar{Acknowledgment}
R. Jackiw thanks C. Kiefer and
K. Kucha\v{r} for discussions about their work.


\begin{thebibliography}{99}
\bibitem{ref:14a}
In fact, $q_{(0)}$ may also have a component along~$I$. Such
generalization would lead to a new and two-dimension specific
interaction~\cite{ref:9,ref:4}, which we do not consider in this paper.
\bibitem{ref:14b}
A coupling of the form $\sqrt{-g}$ $\varphi_3 R$
can also be
incorporated in our formalism by adding
${\cal A} \epsilon^{\mu\nu} \left( \case{1}{2} q^a \epsilon_{a b}
\partial_\mu q^b - \langle q \mid A_\mu \rangle \right) \partial_\nu
\varphi_3$
to the matter Lagrangian~\cite{ref:9,ref:4}.
Such a term is not gauge invariant but its
variation is a total derivative, hence the resulting action is
gauge invariant. We do not include this interaction in the present
discussion.
\bibitem{ref:15a}
Note that the functional derivative with respect to $X^{\pm}\/$ may be
written in terms of one with respect to $X^{\pm \prime} :
\frac{\delta}{\delta X^{\pm} (\sigma)} = - \partial_{\sigma}
\frac{\delta}{\delta X^{\pm \prime} (\sigma)}\/$.  Therefore $X^{\pm
\prime} (\sigma)\/$ is the only gravity variable in~(\ref{eq:4:12b}).
\end{thebibliography}
\end{document}